\documentclass[11pt]{article}
\pdfoutput=1
\usepackage{jheppub}
\usepackage{amssymb,amsmath}
\usepackage{graphicx}
\usepackage{subcaption}
\usepackage{hyperref}
\usepackage{bm}
\usepackage{soul}

\usepackage{xcolor}
\newcommand{\di}[1]{\textcolor{blue}{\bf #1}}

\title{Suppression exponent for multiparticle production in~$\lambda\phi^{4}$ 
  theory}
\author[a,b]{S.V. Demidov}
\author[a,b,c]{B.R. Farkhtdinov}
\author[a,d]{D.G. Levkov}
\affiliation[a]{Institute for Nuclear Research of the Russian Academy of Sciences, 
  Moscow 117312, Russia}
\affiliation[b]{Moscow Institute of Physics and Technology, 
  Dolgoprudny 141700, Russia}
\affiliation[c]{I.M. Sechenov First Moscow State Medical University,
  Moscow 119991, Russia}	
\affiliation[d]{Institute for Theoretical and Mathematical
  Physics, MSU, Moscow 119991, Russia}

\abstract{
  We compute the probability of producing $n$ particles from few
    colliding particles in the unbroken 
  $(3+1)$-dimensional~$\lambda\phi^4$ theory. To this end we
  numerically implement   semiclassical method of singular solutions
  which works at~${n \gg 1}$ in the weakly coupled regime~${\lambda \ll 
    1}$. For the first time, we obtain reliable 
  results in the region of exceptionally large final-state
  multiplicities ${n\gg \lambda^{-1}}$ where the probability decreases
  exponentially with~$n$, ${{\cal P}(\mbox{few} \to n) \sim
    \exp\{f_\infty(\varepsilon) \, n\}}$, and its slope
  $f_{\infty}< 0$ depends  on the mean kinetic energy~$\varepsilon$
  of produced particles. In the opposite case~${n\ll 
    \lambda^{-1}}$ our data match well-known tree-level result, and
  they interpolate between the two limits at~$n \sim
  \lambda^{-1}$. Overall, this proves exponential suppression of the
  multiparticle production probability at~${n\gg 1}$ and arbitrary~$\varepsilon$
  in the unbroken theory. Using numerical solutions, we critically
  analyze  the mechanism for multiple Higgs boson production suggested
  in the literature. Application of our technique to the scalar theory with
  spontaneously broken symmetry can eradicate (or confirm) it  in the
  nearest future.
}
\dedicated{In memory of Valery Rubakov}
\preprint{INR-TH-2022-026}

\begin{document}
\maketitle

\section{Introduction and main results}
\label{sec:intro}
\begin{figure}
  \centerline{\includegraphics{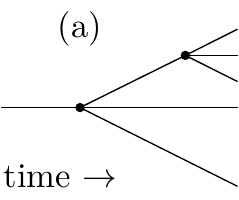}\hspace{15mm}
  \includegraphics{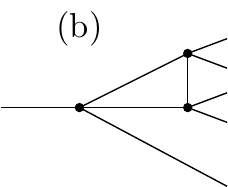}\hspace{15mm}
  \includegraphics{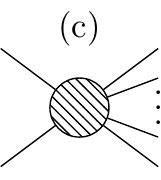}\hspace{15mm}
  \includegraphics{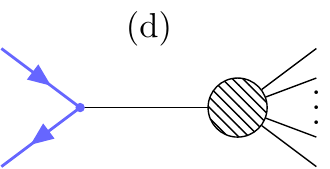}}

  \vspace{2mm}
  \caption{Examples of (a) tree  and (b) one-loop diagrams describing creation
    of~$n=5$ bosons from one off-shell particle in the scalar
    $\lambda \phi^4$ theory. The numbers of such graphs grow with the  
     final-state multiplicity as~$\sim n!$ and~$\sim n^2n!$, respectively. (c),  
    (d)~Processes $\mbox{few}\to n$ for different
    operators $\hat{\cal O}$ in Eq.~(\ref{eq:1.1}).}
  \label{fig:dia}
\end{figure}

Perturbative method is efficient for computing few-to-few scattering
amplitudes in weakly coupled field theories. But it may become
unreliable~\cite{Ringwald:1989ee, Espinosa:1989qn, Cornwall:1990hh,
  Goldberg:1990qk} if the number~$n$ of particles in the final
state exceeds the inverse coupling constant~$\lambda^{-1}$ of the
theory. Indeed, the count of tree diagrams contributing to $\mbox{few} \to n$ 
processes in models of scalar fields grows
factorially~\cite{Cornwall:1990hh, Goldberg:1990qk, Brown:1992ay}
with~$n$, and $l$-loop corrections add even more~--- of order
$n!n^{l+1}$~--- terms, see~\cite{Voloshin:1992nu, 
  Smith:1992rq,   Libanov:1994ug} and Figs.~\ref{fig:dia}a,b. As
a consequence, perturbative series for  
the respective amplitudes are proportional to  $n!$, go
in powers of~$\lambda n$ instead of $\lambda$, and explode
at~${n\gtrsim  \lambda^{-1}}$~\cite{Libanov:1994ug, Libanov:1997nt,
  Dine:2020ybn}. This simplified bookkeeping is supported by explicit
calculations in the scalar field theories at tree~\cite{Brown:1992ay,
  Argyres:1992np, Smith:1992kz, Libanov:1993qf,  Khoze:2014zha} and 
one-loop~\cite{Voloshin:1992nu, Smith:1992rq, Schenk:2021yea,
  Khoze:2022fbf} levels, both at the mass threshold of $n$ final
particles and for their nonzero spatial momenta~\cite{Argyres:1992kt,
  Libanov:1994ug, Khoze:2014kka}. It also agrees with
the intuition acquired from one-dimensional quantum mechanics~\cite{Bachas:1991fd,
  Jaeckel:2018ipq, Jaeckel:2018tdj}.

Later, it was observed~\cite{Libanov:1994ug} that the parts of
perturbative series going in powers of~$\lambda n$ can be  resummed
into an exponent of a universal ``holy  grail''
function~$F$. Consider, say, the scalar~$\lambda\phi^4$
theory and inclusive probability  ${\cal P}_n(E)$ of producing $n$
scalar quanta with total energy~$E$ from the few-particle
initial state~$\hat{\cal O} | 0\rangle$. At~${n \sim 
  \lambda^{-1} \gg 1}$ this probability is expected to have the
form~\cite{Libanov:1994ug}, 
\begin{equation}
\label{eq:1.1}
{\cal P}_n ( E)  \equiv \sum_{f} |\langle
f; E,n|\hat{\cal S}\,\hat{\cal O}|0\rangle|^2 \sim {\rm
 e}^{F(\lambda n, \, \varepsilon)/\lambda}\;,
\end{equation}
where the sum covers all final states with given $n$ and $E$,
$\hat{\cal S}$ is the S-matrix, and the exponent in the right-hand
side collects all leading terms of the perturbative series at
  fixed~$\lambda n$. The initial state operator $\hat{\cal O}$   
may create two $\phi$-quanta~--- then~${\cal P}_n$ is proportional
  to the familiar~${2\to n}$ cross section (Fig.~\ref{fig:dia}c).
Otherwise, one can take~${\hat{\cal  O}\propto \hat{\phi}(0)}$ if
  the off-shell $\phi$-boson is initially produced in an
  external 
  collision, see Fig.~\ref{fig:dia}d.  In any case, the exponent~$F$
  is conjectured to be {\it universal}~\cite{Libanov:1995gh},
  i.e.\ independent of the operator~$\hat{\cal O}$ as long as the
  latter creates~${\ll \lambda^{-1}}$ particles from the vacuum. This
  makes $F$  a function of two variables: the
  combination~${\lambda n  
    \sim O(1)}$ and mean kinetic  energy~${\varepsilon \equiv E/n -
    m}$ of final particles with mass $m$.

To date, the form (\ref{eq:1.1}) of the probability  and
universality of the exponent are confirmed in the~$\lambda
\phi^4$ theory~\cite{Libanov:1994ug, Libanov:1995gh}   for the
    two leading terms of $F$ expansion in~$\lambda n$, for many
  expansion orders in the  analogous quantum
  mechanics~\cite{Jaeckel:2018ipq, Jaeckel:2018tdj}, and for the
  sister processes of underbarrier tunneling between the 
few-particle and multiparticle states\footnote{In the latter case
universality of the exponent is called Rubakov-Son-Tinyakov
  conjecture~\cite{Rubakov:1992ec}.}~\cite{Tinyakov:1991fn,
  Mueller:1992sc, Bonini:1999kj, Levkov:2008csa}. All these tests 
are nonperturbative because the right-hand side of
  Eq.~(\ref{eq:1.1}) includes arbitrarily high powers of~$\lambda$ 
even in the simplest case when~$F$ is cropped to~$O(\lambda
  n)$ and~$O(\lambda  n)^2$ terms. At  the same time, no reliable  
first-principle calculation 
of the exponent has yet been performed at~${n \sim \lambda^{-1}}$
  in any field theoretical model, see
  Refs.~\cite{Kiselev:1992hk, Rubakov:1992gi, Kuznetsov:1997az,
    Bezrukov:2003er, Bezrukov:2003qm, Levkov:2004tf, 
    Levkov:2004ij, Demidov:2011dk, Demidov:2015nea, Demidov:2015bua}
  for similar results in the case of tunneling.

Expression~(\ref{eq:1.1}) reveals exponential sensitivity of the
scattering probability to the number of particles in the final
state. One asks~\cite{Ringwald:1989ee, Espinosa:1989qn} whether
it
may become unsuppressed at sufficiently large~${n \sim
  \lambda^{-1}}$. Recently, this question was acutely posed 
in the context of the so-called ``Higgsplosion'' scenario~\cite{Khoze:2017tjt, 
  Khoze:2017ifq, Khoze:2018mey}:  the exponent for
producing~$n$ nonrelativistic Higgs bosons from  two colliding gluons
was suggested to have the form, 
\begin{equation}
  \label{eq:1}
  F_{\mathrm{Higgsplosion}} \approx\lambda n\,  \ln\frac{\lambda n}{4}
  + \frac32 \lambda n \, \ln \frac{\varepsilon}{3 \pi m} +
  \frac{\lambda n}{2} + 0.854\, (\lambda n)^{3/2} \qquad \mbox{at}
  \quad n \leq n_*\;.
\end{equation}
Here $m$, $\lambda$, and $\varepsilon \ll m$ are the mass, quartic
constant, and mean kinetic energy of the final-state Higgs bosons, $E = n
(m+\varepsilon)$ is the collision energy, and $n_*$ is defined by
$F_{\mathrm{Higgsplosion}} (\lambda n_*) = 0$. Importantly,  Eq.~(\ref{eq:1}) 
was derived semiclassically at $n \sim \lambda^{-1}$, albeit with
daring assumptions on the structure of semiclassical
solutions~\cite{Khoze:2017ifq, Khoze:2018mey}. It may well be valid in 
the entire region of multiplicities~${n\leq n_*  \sim 3.08\,
  \lambda^{-1} \ln^2 (\varepsilon/m)}$ where the exponent is
  non-positive~--- then the transitions become
unsuppressed at $n \approx n_*$. At 
larger $n$  corrections to Eq.~(\ref{eq:1}) should
prevail~\cite{Khoze:2017tjt, Belyaev:2018mtd} and unitarize the theory
because~the probability ${{\cal   P}_n }$ cannot be exponentially large\footnote{Also, large-$n$ asymptotic of   Eq.~\eqref{eq:1} is 
  inconsistent with locality of quantum theory~\cite{Monin:2018cbi,
    Khoze:2018qhz}.}, cf.~\cite{Zakharov:1991rp, Veneziano:1992rp,
  Maggiore:1991vi}. But even in its limited parameter space
Eq.~(\ref{eq:1}) can drastically change the entire Higgs phenomenology
predicting explosive production of these particles in high-energy 
collisions and in decays of new heavy
resonances~\cite{Khoze:2017tjt,   
  Khoze:2017lft}, cf.~\cite{Voloshin:2017flq}.  

On the other hand, one finds this outstanding possibility challenging from
  the consistency side of quantum theory. Indeed, recall that the
inclusive probability of high-energy scattering is related~--- by
optical theorem and dispersion  relations~--- to the Green's function
of few field operators at low momenta. If the transitions from
  ``few'' to ``many'' were unsuppressed at high energies, the Green's  
functions would receive sizeable contributions from 
virtual multiparticle states, and that would break perturbative expansion at  
small momenta~\cite{Zakharov:1991rp,    Libanov:1997nt}. This
argument may be brushed off as inconclusive~\cite{Khoze:2018qhz},
but it certainly raises the stakes: either Eq.~(\ref{eq:1}) is invalid
or one  of the building blocks of a consistent quantum field theory~---
dispersion  relations or the perturbative method~--- should be
abandoned. We will return to this issue in the Discussion section. 

Another warning comes from simulations of classical waves. If the
probability of multiparticle production were of order one, the time-reversed
processes, namely, conversion of many particles into 
few highly energetic quanta would also proceed
classically~\cite{Rubakov:1995hq, Rebbi:1996zx, Demidov:2011eu}. The
latter conversion was not, however, observed in 
evolutions of classical wave packets~--- read, collisions of
multiparticle states~--- despite Monte Carlo optimization over the available
parameter space~\cite{Demidov:2018czx}. A possible loophole here is a
different model: unbroken~$\lambda\phi^4$ theory in
Ref.~\cite{Demidov:2018czx} as compared to the spontaneously 
  broken case used for deriving Eq.~(\ref{eq:1}).

In this paper we numerically compute the exponent $F(\lambda n,\,
\varepsilon)$ at arbitrary~$\lambda n$ from first 
principles in the scalar field theory. Up to our knowledge, no
  calculation of this kind was performed before, see Ref.~\cite{Demidov:2021rjp} for the
accompanying work. We exploit the 
same  semiclassical method~\cite{Son:1995wz} as in the 
  studies on ``Higgsplosion''~\cite{Khoze:2017ifq, Khoze:2018mey},
but do not make additional assumptions on the structure of saddle-point
solutions. We consider $(3+1)$-dimensional $\lambda \phi^4$
theory with positive mass term $m^2 > 0$ and no 
spontaneous symmetry breaking:
\begin{equation}
\label{eq:2.1}
S = \frac{1}{2\lambda} \int d^4 x \,  \left( - \phi \Box
\phi - m^2 \phi^2 - \phi^4/2\right)\,,
\end{equation}
where the coupling~${\lambda \ll 1}$ appears in front of the action and
hence governs loop expansion; one can bring it in front of
  the~$\phi^4$ term using the field redefinition~$\phi \to 
  \phi \sqrt{\lambda}$.

We rely on the semiclassical technique of D.T.~Son~\cite{Son:1995wz}
developed as an adaptation of L.D.~Landau method for calculating
matrix elements in quantum mechanics~\cite{Landau:1991wop};
see also~\cite{Voloshin:1990mz, 
  Khlebnikov:1992af, Gorsky:1993ix, Diakonov:1993ha}. It applies at $n \gg 1$ and $\lambda \ll 1$ and is based
on the universality of the exponent in Eq.~(\ref{eq:1.1}). Namely,
since~$F$ is independent of the  few-particle operator $\hat{\cal O}$,
we can take it in the form 
\begin{equation}
\label{eq:2.2}
\hat{\cal O} = \exp\left\{-\frac1\lambda \int d^3\boldsymbol{x} \,
J(\boldsymbol{x})\, \hat{\phi}(0,\, \boldsymbol{x})\right\}
\end{equation}
that describes a classical source $J(\boldsymbol{x})$ acting
at~${t=0}$. The latter source creates~$O(J^2 / \lambda)$
particles from the vacuum, i.e.\ a few-particle initial
state with  multiplicity~${\ll \lambda^{-1}}$ at~${J \ll  
  O(\lambda^0)}$. Then the universality conjecture guarantees
that the limit 
\begin{equation}
  \label{eq:3}
  F(\lambda n,\varepsilon) = \lim_{J \to 0} F_J(\lambda n , \, \varepsilon)
\end{equation}
exists and is independent of the source profile, where~$F_J$ in the
right-hand side is computed at nonzero $J(\boldsymbol{x})$. 

\begin{sloppy}

The above observation opens up a way to the semiclassical description
because at~${O(J^2/\lambda) \gg 1}$ and~${n\gg1}$ both the initial
  and final states of the process include many particles. Therefore,
  one can compute $F_J$ semiclassically and then take the
  limit~$J\to 0$, thus arriving at
the exponent of the~$\mbox{few} \to n$ production probability. We
stress that the last limit  brings 
the initial multiplicity to the  region ${1 \ll O(J^2/  \lambda)
  \ll \lambda^{-1}}$ where the exponent is already universal but the
semiclassical method is still applicable. 

\end{sloppy}

At finite $J$, the standard semiclassical machinery~\cite{Son:1995wz, 
  Rubakov:1992ec, Tinyakov:1992dr} works in the following
  way. One writes the probability~(\ref{eq:1.1}) in the form of a
path integral and evaluates  the latter in the saddle-point
approximation. The respective saddle-point
configurations~$\phi_{\mathrm{cl}} 
(t,\,\boldsymbol{x})$ are generically complex. They satisfy the
classical field  equation in the presence of an external
source~$J(\boldsymbol{x})$ and the boundary conditions at~${t\to \pm \infty}$
depending on $\varepsilon$ and~$\lambda n$. Once the semiclassical solutions are found,~$F_J$ 
can be computed using the value of the action functional
on~$\phi_{\mathrm{cl}}(t,\,\boldsymbol{x})$. Notably, the
semiclassical configurations become singular in the limit~${J \to
  0}$;  that is why the overall technique is called
the method of singular solutions~\cite{Son:1995wz}. Previously, it was
shown~\cite{Son:1995wz, Libanov:1997nt} that this method correctly  
reproduces tree-level and one-loop multiparticle amplitudes at the threshold
($\varepsilon=0$) in the $\lambda\phi^4$ theory. Besides,
in Ref.~\cite{Bezrukov:1998mei} it was used to calculate tree-level
suppression exponent  at $\lambda n \ll 1$ and
arbitrary~$\varepsilon$. But apart from the controversial
   ``Higgsplosion'' studies, the most interesting case~$n\gtrsim
\lambda^{-1}$ was never considered.  

In this paper we numerically find the saddle-point solutions
$\phi_{\mathrm{cl}}(t,\, \boldsymbol{x})$ at arbitrary~$\lambda n $
and~$\varepsilon$ in the model (\ref{eq:2.1}). We make no assumptions
on their properties or analytic structure. We reliably
select physically relevant configurations that give dominant
contributions to the probability. Namely, at certain nonzero~$J$ the
out-particles are  mainly produced  by the  classical source itself,
while the interaction of the scalar field can be ignored. In this
case the physical solutions can be found in the free theory with a
source. Switching on interaction and  gradually decreasing~$J$ to zero
at fixed $\lambda n$ and $\varepsilon$, we arrive at a continuous
branch of relevant saddle-point configurations. The subsequent
extrapolation~${J\to 0}$ gives the sought-for singular solutions and their 
suppression exponent~$F$ in the the broad range of~${n\gg   1}$ 
and~$\varepsilon$.

\begin{figure}
  \centerline{\includegraphics{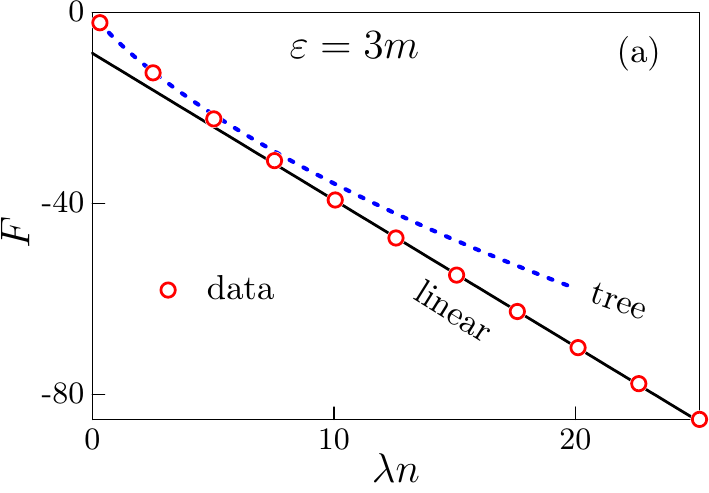}\hspace{5mm}
    \unitlength=1mm
    \begin{picture}(72,50)
      \put(0,0){\includegraphics{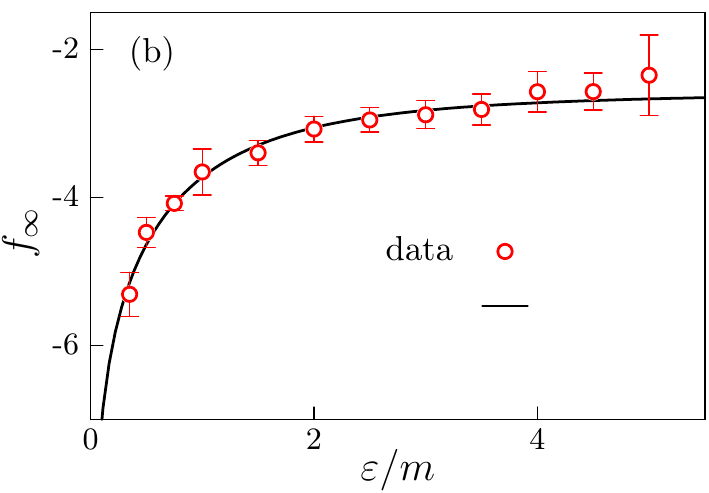}}
      \put(34.4,18){\small fit~\eqref{eq:4.15}}
    \end{picture}}
  \caption{(a) The exponent $F(\lambda n,\, \varepsilon)$ of
     the multiparticle production probability in the
     model~(\ref{eq:2.1}) at~${\varepsilon = 3 m}$. Numerical data 
     (circles) interpolate between the tree-level result at~${\lambda n
     \ll 1}$ (dashed line and Eq.~\eqref{eq:4.1} from the main
     text) and  linear asymptotic~\eqref{eq:4.9} at~${\lambda n \gg 1}$  
     (solid line). (b)~The slope~$f_\infty (\varepsilon)$ of the exponent
     at large~$\lambda n$ as a function of~$\varepsilon$.}
   \label{fig:intro_results}
\end{figure}

Our result for the exponent $F(\lambda n, \, \varepsilon)$
is demonstrated in Fig.~\ref{fig:intro_results}a (circles) at the
  exemplary value $\varepsilon = 3m$ with numerical accuracy better than the
circle size. We see that~$F$ monotonically decreases with $\lambda
n$. As expected, at $\lambda n \ll 1$ it coincides with the 
contribution of tree-level diagrams~\cite{Bezrukov:1995ta} (dashed
line), see also~\cite{Demidov:2021rjp}. In 
the opposite limit $\lambda n \gg 1$ our numerical data are well
fitted by the linear function (solid line in the figure):  
\begin{equation}
  \label{eq:4.9}
F \to  \lambda n f_{\infty}(\varepsilon) + g_{\infty}(\varepsilon)
\qquad \mbox{or} \qquad {\cal P}_n \to \mathrm{e}^{n
  f_{\infty}(\varepsilon) + g_{\infty}(\varepsilon)/\lambda}
\qquad \mbox{at}\qquad \lambda n \to +\infty\,,
\end{equation}
where $f_{\infty}$ and $g_{\infty}$ are negative. In the main text we will
show that results at other $\varepsilon$ have similar qualitative
behavior, although $f_{\infty}$ and $g_{\infty}$ depend
  on$\varepsilon$. In particular, Fig.~\ref{fig:intro_results}b 
demonstrates the slope~$f_\infty (\varepsilon) < 0$
increasing with energy. It can be approximated by the
expression (solid line in the figure) 
  \begin{equation}
  \label{eq:4.15}
  f_{\infty}(\varepsilon) \approx -\frac{3}{4}\ln \left[ (d_{1}
    m/ \varepsilon)^{2}  + d_2\right]\,, \qquad  \quad
  d_i \approx \{10.7, \,  30.7 \}
\end{equation} 
 that has physically motivated asymptotics at $\varepsilon \to 0,\,
+\infty$; see the main text for derivation. Minimal slope $f_{\infty}
  \to -2.57 \pm 0.06$  is achieved in the ultrarelativistic limit
  $\varepsilon \to +\infty$.

\begin{figure}
  \centerline{\unitlength=1mm
    \begin{picture}(72,50)
      \put(0,0){\includegraphics{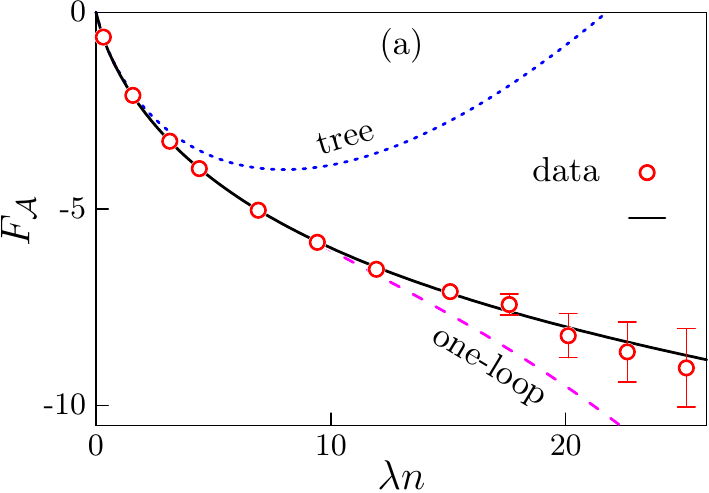}}
      \put(47.7,27){\small fit (\ref{eq:4.5a})}
    \end{picture}\hspace{7mm}
  \includegraphics{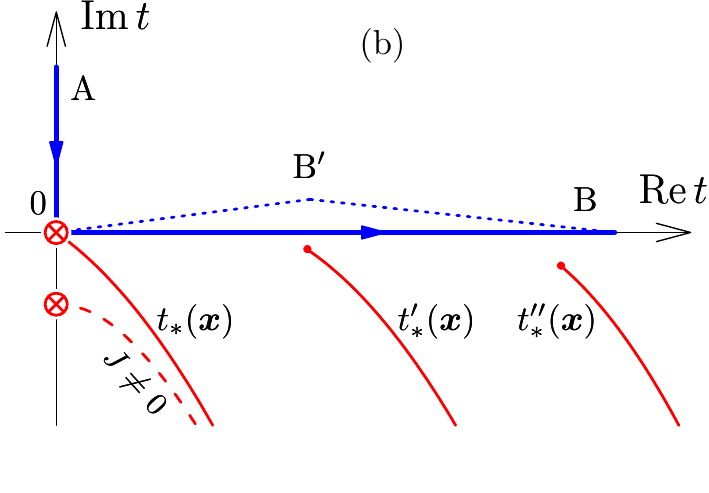}}
  \caption{(a) Suppression exponent $F_{\cal A} (\lambda n)$ of
      the amplitude to produce $n$ particles at the mass threshold,
    see\  Eq.~(\ref{eq:4}). (b)~Complex time contour A0B for the
    semiclassical boundary value problem (thick solid line) and the
    singularities $t_*(\boldsymbol{x})$, $t_*'(\boldsymbol{x})$,
      $t_{*}''(\boldsymbol{x})$ of the semiclassical 
    solutions (thin lines starting from crossed circles or filled points);
    not to scale.}
  \label{fig_cont}
\end{figure}

It is remarkable that the probability~(\ref{eq:1.1}) can be used to
calculate the amplitude ${\cal A}_n$ of producing~$n$ particles at the
mass threshold. Indeed, in the limit $\varepsilon \to 0$ 
 a single out-state remains: the one with zero spatial
 momenta of all final 
particles. The amplitude of transition to this state is determined
by the ratio of the inclusive probability to the
$n$-particle phase volume~${\cal   V}_n(\varepsilon)/n!$
at $\varepsilon\to 0$:
\begin{equation}
  \label{eq:4}
  |{\cal A}_n|^2 \sim \lim_{\varepsilon \to 0} \frac{n!}{{\cal V}_n}\;
  \mathrm{e}^{F/\lambda} \,\sim  n! \,m^{4-2n}\, \mathrm{e}^{2F_{\cal
      A}(\lambda n)/\lambda}\,,
\end{equation}
where the factor $n!$ explicitly accounts for particle
identity, the standard expression for~${\cal V}_n$ is given in
the main  text, and the last equality fixes the expected
parametric form of the amplitude at~${n \sim
  \lambda^{-1}}$. Extrapolating numerical results to $\varepsilon=0$,
we  get the exponent~$F_{\cal A} (\lambda n)$ which is displayed by
circles with errorbars in Fig.~\ref{fig_cont}a. At small $\lambda   
n$ these data are close  to the tree-level exponent
of Ref.~\cite{Brown:1992ay} (dotted line) and even closer to the one-loop
result of Refs.~\cite{Voloshin:1992nu,   Libanov:1994ug} (dashed line). At large 
$\lambda n$ the data deviate from the perturbative results, but
they are well fitted  by the function with linear asymptotic 
as~$\lambda n \to +\infty$ (solid line and
  Eq.~(\ref{eq:4.5a}) from the main text). In the latter
large-$\lambda n$ region the 
amplitude equals
\begin{equation}
  \label{eq:5}
  |{\cal A}_n| \sim m^{2-n}\sqrt{n!} \; \mathrm{e}^{ nf_\infty' +
    g_\infty' / \lambda}\,, \;\;
  f_{\infty}' = -0.062 \pm 0.026\,, \;\;
  g_{\infty}' = -9.7 \pm 1.2\;\;\;\mbox{at} \;\;\; n\gg \lambda^{-1}\,.
\end{equation}
We will explain below that Eq.~(\ref{eq:5}) does not contradict to
unitarity of quantum theory despite the factorial dependence on $n$.

In a nutshell, our results prove that the probability of producing
$n\gg 1$ particles from few colliding quanta is exponentially
suppressed in the unbroken $\lambda \phi^4$ theory. In addition to
already presented data, below we visualize the suppression
exponent at different~$\varepsilon$, fit it with a convenient
formula at finite  $\lambda n$, and provide tabulated raw data in the
ancillary files~\cite{ancillary}. We also confirm universality of
the exponent by comparing results at different  source
profiles $J(\boldsymbol{x})$. Finally, we will discuss
self-consistency of Eqs.~\eqref{eq:4.9} and~(\ref{eq:5}), physical
constraints on $F$, and reiterate interconnections between the 
multiparticle probability, unitarity, perturbative series, and
dispersion relations. 

An important part of our study is the analysis of 
configurations $\phi_{\mathrm{cl}}(t,\, \boldsymbol{x})$ saturating
the probability~(\ref{eq:1.1}). Although being complex and with no immediate 
physical meaning, they characterize the dynamics of the process and
specify its most probable final state. Studying 
the analytic structure of the solutions, we find out that one of
the assumptions underlying the  
``Higgsplosion''  scenario~\cite{Khoze:2017ifq, Khoze:2018mey} is not
generic. This puts Eq.~(\ref{eq:1}) on shaky ground despite the fact
that it was obtained in a different model.

The rest of this paper is organized as follows. We review the
semiclassical method of singular solutions and formulate it on the
lattice in
Sec.~\ref{sec:singul_solut}. Section~\ref{sec:way_to_sing_solut}
  explains our techniques for selecting physically relevant 
  solutions and for extrapolating the data to~${J \to 0}$. We
    present results in Sec.~\ref{sec:numerical-results} and
    discuss them in 
  Sec.~\ref{sec:conclus}. Appendices~A,B, and~C provide
  details of numerical methods, saddle-point solutions in the linear
  theory, and analysis of singularities, respectively. 
 
\section{Semiclassical method}
\label{sec:singul_solut}

\subsection{Saddle-point equations}
\label{sec:saddle-point-equat}

We start by reviewing equations for the saddle-point
configurations $\phi_{\mathrm{cl}}(t,\,\boldsymbol{x})$ 
saturating
path integral for the probability (\ref{eq:1.1}) at~$\lambda \ll 1$
and large~$n$, see~\cite{Son:1995wz, Rubakov:1992ec, Tinyakov:1992dr,
  Khoze:2018mey} for derivation.

Generically, $\phi_{\mathrm{cl}}$  are complex and satisfy the 
field equation
\begin{equation}
\label{eq:2.4}
\Box \phi_{\mathrm{cl}} +
m^2\phi_{\mathrm{cl}} + \phi^3_{\mathrm{cl}} = i
J(\boldsymbol{x})\, \delta(t)\,,
\end{equation}
where the source in the right-hand side originates from the operator
$\hat{\cal O}$ in the form~\eqref{eq:2.2}. Boundary conditions for
this equation are related to the initial and final states of the
process.

It will be convenient to analytically continue
$\phi_{\mathrm{cl}}(t,\, \boldsymbol{x})$ onto the complex time
contour~A0B in Fig.~\ref{fig_cont}b, i.e.\ perform a partial Wick
rotation. Then the first boundary condition requires the field to vanish
in the infinite past along the Euclidean time axis:  
\begin{equation}
  \label{eq:2.5}
  \phi_{\mathrm{cl}}(t,\, \boldsymbol{x}) \to 0 \qquad \mbox{as} \qquad t \to
  +i\infty\,.
\end{equation}
This corresponds to initial vacuum in Eq.~(\ref{eq:1.1}). In the
infinite future $t\to +\infty$, the semiclassical solution 
describes a state of $n$ free particles. Hence, it  should 
linearize into a superposition of free waves: 
\begin{equation}
  \label{eq:2.6}
  \phi_{\mathrm{cl}} (t,\, \boldsymbol{x}) \to  \int \frac{d^3\boldsymbol{k}\;  {\rm
      e}^{i{\boldsymbol{kx}}}}{(2\pi)^{3/2}\sqrt{2\omega_{\boldsymbol{k}}}}
  \,\left[\, a_{\boldsymbol{k}}{\rm e}^{-i\omega_{\boldsymbol{k}}t}  
  + b_{-\boldsymbol{k}}^*{\rm e}^{i\omega_{\boldsymbol{k}}t}\, \right] \qquad
  \mbox{as} \qquad t\to +\infty\,,
\end{equation}
where $\omega_{\boldsymbol{k}} \equiv \sqrt{\boldsymbol{k}^2 + m^2}$
and $a_{\boldsymbol{k}}$ and $b_{\boldsymbol{k}}$ are the negative-
and positive-frequency amplitudes, respectively. The second boundary condition
relates the amplitudes:
\begin{equation}
\label{eq:2.7}
a_{\boldsymbol{k}} = {\rm e}^{-\theta + 2\omega_{\boldsymbol{k}}T}\, b_{\boldsymbol{k}}\,.
\end{equation}
One can show~\cite{Rubakov:1992ec} that this equation corresponds to
inclusive final state with given energy and multiplicity in
Eq.~(\ref{eq:1.1}). 
Parameters~$T$ and~$\theta$ in its
right-hand side are the Lagrange multipliers
related to~$E$  and~$n$ via the standard expressions
\begin{equation} 
\label{eq:2.8}
\lambda E = \int d^3 \boldsymbol{ k}\, \omega_{\boldsymbol{k}}\,
a_{\boldsymbol{k}} b_{\boldsymbol{k}}^* 
\,, \quad\qquad  \lambda n = \int d^3 {\boldsymbol{k}}\,
 a_{\boldsymbol{k}}b_{\boldsymbol{k}}^*\,. 
\end{equation}
We will also use kinetic energy per final particle $\varepsilon
\equiv E/n - m$.

It is worth noting that a complete Wick rotation to the Euclidean axis
cannot be performed. Indeed, $b_{\boldsymbol{k}}$ cannot vanish for
all $\boldsymbol{k}$ at 
nonzero~$E$ and~$n$ due to Eqs.~(\ref{eq:2.8}). Thus, the
positive-frequency
part of the solution (\ref{eq:2.6}) grows exponentially at~$t \to
-i\infty$, ruins linearity and makes it impossible to impose free-wave
boundary  conditions in that region. On the other hand, the
semiclassical equations can be consistently formulated on the
contour~A0B, and the source $J(\boldsymbol{x})\delta(t)$  can be
placed right in its corner  at~$t=0$. The latter fact is made
  explicit by integrating Eq.~(\ref{eq:2.4}) across~$t=0$. We get,
\begin{equation}
  \label{eq:6}  
  \partial_t \phi_{\mathrm{cl}}(+0,\, \boldsymbol{x})  -
  \partial_t\phi_{\mathrm{cl}}(+i0,\, \boldsymbol{x})  =
  iJ(\boldsymbol{x})\,, \qquad 
  \phi_{\mathrm{cl}}(+0 ,\, \boldsymbol{x}) =
  \phi_{\mathrm{cl}}(+i0,\, \boldsymbol{x}) \qquad \mbox{at} \quad t=0\,.
\end{equation}
Hence, one can find the solutions of Eq.~(\ref{eq:2.4})
with zero right-hand side on the parts A0 and~0B of the time
contour and then glue them at $t=0$ using Eq.~\eqref{eq:6}.

Equations~\eqref{eq:2.4}--\eqref{eq:2.8} form a complete boundary
value problem for the 
semiclassical configuration 
$\phi_{\mathrm{cl}}(t, \, \boldsymbol{x})$ and Lagrange multipliers
$T$, $\theta$. Once the equations are solved, one calculates the
suppression exponent~\cite{Son:1995wz}, 
\begin{equation}
\label{eq:2.9a}
F_J = 2\lambda ET - \lambda n\theta-2 \lambda\mathrm{Im}\,
  S[\phi_{\mathrm{cl}}]
- 2\mathrm{Re} \int d^3 \boldsymbol{x} \, J(\boldsymbol{x}) \,
\phi_{\mathrm{cl}} (0,\, \boldsymbol{x})\,,
\end{equation}
where the first two terms come from the
non-vacuum final state, the classical action~(\ref{eq:2.1}) in
  the third term is evaluated on $\phi_{\mathrm{cl}}$, and the last
term accounts for the 
insertion of the operator~$\hat{\cal O}$. Note that the semiclassical
equations involve $\lambda$, $n$, and $E$ only in the
combinations~$\lambda n$ and~$\lambda E$, see Eq.~(\ref{eq:2.8}). This
makes the rescaled classical action $\lambda S[\phi_{\mathrm{cl}}]$
and the semiclassical exponent~$F_J$ depend on two parameters:
$\lambda n$ and $\varepsilon \equiv E/n - m$. 

It is also worth pointing out that the Lagrange multipliers~$T$
and~$\theta$ automatically satisfy Legendre transform
relations~\cite{Son:1995wz}:
\begin{equation}
\label{eq:2.11}
2T = \frac{\partial F_J}{ \partial (\lambda E)} \,, \qquad \qquad
\theta = -\frac{\partial F_J}{\partial(\lambda n)}\,.
\end{equation}
This discloses them as derivatives of the suppression exponent
$F_J(\lambda n, \, \lambda E)$. We will use Eqs.~(\ref{eq:2.11})
as a cross-check of the numerical method and a cheap way to
increase precision.

The last step of the semiclassical method is to send $J \to 0$. Let us
demonstrate~\cite{Son:1995wz} that the semiclassical solutions become
singular in this limit. Consider their energy
\begin{equation}
  \label{eq:7}
  {\cal E}(t) = \frac1{2\lambda}\int d^3 \boldsymbol{x} \left[
  (\partial_t \phi_{\mathrm{cl}})^2 + (\partial_{\boldsymbol{x}}
  \phi_{\mathrm{cl}})^2 + m^2 \phi_{\mathrm{cl}}^2 +  \phi_{\mathrm{cl}}^4/2 \right]\,,
\end{equation}
which separately conserves on the Euclidean and Minkowski parts
A0 and 0B of the time contour. Namely, ${\cal E} = 0$ on
  the part~A0 and ${\cal E} = E$ on the part~0B due to boundary 
  conditions~(\ref{eq:2.5}) and (\ref{eq:2.8}). The energy jumps
at $t=0$ due to presence of the classical source~$J$. We therefore obtain,  
\begin{equation}
  \label{eq:8}
  \lambda E = \lambda {\cal E}(+0) - \lambda{\cal E}(+i0) =
  \frac{i}2 \int d^3 \boldsymbol{x} \, J(\boldsymbol{x})
  \left[\partial_t\phi_{\mathrm{cl}}(+0,\, \boldsymbol{x}) +
    \partial_t\phi_{\mathrm{cl}}(+i0,\, \boldsymbol{x}) \right]\,,
\end{equation}
where Eqs.~(\ref{eq:6}) were  used in the last equality. Now, it is
clear that $\partial_t \phi_{\mathrm{cl}}$ should become singular
at  $t=0$ in the limit $J \to 0$, or the energy $E$ would vanish. 

Another clarification of the analytic structure comes from the
solution at $\lambda n = \lambda E = 0$ and $J=0$ which is known
analytically~\cite{Brown:1992ay,   Son:1995wz}. It
is spatially homogeneous:
\begin{equation} 
  \label{eq:9}
  \phi_{\mathrm{cl}}(t,\, \boldsymbol{x}) = - im\sqrt{2} \big/ \sin(mt
    \mathrm{e}^{i\epsilon'})  \qquad \mbox{at} \qquad  \lambda
    n = \lambda E = 0\,,
\end{equation}
where $\epsilon' \to +0$ is a regulator. One can check that
Eq.~(\ref{eq:9}) solves the field equation with zero source,
  has $a_{\boldsymbol{k}} = 0$, and satisfies the boundary 
conditions~(\ref{eq:2.5}), (\ref{eq:2.6}), and (\ref{eq:2.7}) at~${\theta
  = +\infty}$. Expressions~(\ref{eq:2.8}) then give zero 
  quantum numbers of the final state. We see a singularity
  of~$\phi_{\mathrm{cl}}$   at~${t=0}$~---  or, rather, a flat  
three-dimensional singularity  hyperplane in the four-dimensional 
spacetime. But the configuration~\eqref{eq:9} is also singular at the chain of 
points~${t = \pi k \,  \mathrm{e}^{-i\epsilon'}/m}$ with
  integer~$k$  which reside somewhat below the real time 
  axis at~${k>0}$. Below we   will demonstrate numerically that the spatial
homogeneity of 
solutions is broken at nonzero~$E$ and~$n$, but the qualitative
singularity structure remains. Namely, the singularities 
form~\cite{Son:1995wz, Libanov:1997nt} 
chains of hypersurfaces $t = t_{*}(\boldsymbol{x})$,
$t_{*}'(\boldsymbol{x})$, etc.\ shown in Fig.~\ref{fig_cont}b. The 
first~--- ``main''~--- hypersurface $t_*(\boldsymbol{x})$ passes
the point~${t_* = \boldsymbol{x} = 0}$ at~${J =0}$ and shifts to
$\mathrm{Im}\, t_*<0$ at nonzero source. This corresponds to
singular and regular solutions on the contour A0B{,
  respectively}. We pictured 
the ``main'' singularity  in Fig.~\ref{fig_cont}b by the
  solid ($J=0$) and dashed ($J\ne 0$) lines starting at the
crossed circles.

It is worth noting that the original paper~\cite{Son:1995wz} took one
step forward and tried to derive boundary value problem for the
actual singular solutions at $J =0$. We will not use such
reformulation, as it is inconvenient for numerical implementation.

To sum up, our semiclassical method consists of solving
Eqs.~(\ref{eq:2.4})~---~(\ref{eq:2.8}) and evaluating the
exponent~(\ref{eq:2.9a}). The last step is extrapolation of results
to~${J \to 0}$ according to Eq.~\eqref{eq:3}, which will be also done
numerically.

\subsection{Numerical implementation}
\label{subsecsec:num_implement}

Now, we reformulate the semiclassical boundary value problem on the
lattice. We switch to dimensionless units with $m=1$ and consider
a Gaussian source
\begin{equation}
  \label{eq:2.3}
  J(\boldsymbol{ x}) = j_0 \; {\rm e}^{-\boldsymbol{ x}^2/2\sigma^2}
\end{equation}
of strength $j_0$ and width $\sigma$. Eventually, we will
send~$j_0 \to 0$ and $\sigma \to 0$ at a fixed~$j_0/\sigma$. This
  will make the source local in space and small in amplitude,
i.e.\ similar to the vanishing delta function~${J \to
  (2\pi)^{3/2} \,   j_0 \sigma^3  \,   \delta^{(3)}
  (\boldsymbol{x})}$ used in~\cite{Son:1995wz}. Comparing
  results at different~$j_0 / \sigma$, we will test universality of 
the semiclassical exponent.  

We assume spherical symmetry of the saddle-point configurations:
${\phi_{\mathrm{cl}} = \phi_{\mathrm{cl}}(t, \, r)}$, where ${r \equiv  
  |\boldsymbol{x}|}$. This Ansatz passes the saddle-point
equations~\eqref{eq:2.4}~---~\eqref{eq:2.8} and agrees with all
previously known semiclassical solutions~\cite{Son:1995wz, Bezrukov:1998mei,
  Libanov:1997nt}. Spherical symmetry also complies 
with  insensitivity of the semiclassical exponent to the few-particle
initial state: taking the latter isotropic, one
  can make the entire process rotationally invariant. On
the other hand, the spherical Ansatz leaves only two coordinates $t$
and $r$ and hence  significantly simplifies numerical calculations.

We introduce temporal and spatial lattices with sites~$t_j$  and~$r_i$
covering the complex contour\footnote{\label{fn:1}At large $\lambda n$ the
  chain singularities of the solutions approach the real time axis and
  inflate numerical errors. In that case we deform the
``Minkowskian''  part of the time contour to the line~0B$^\prime$B
shown
in Fig.~\ref{fig_cont}b (dotted).}~A0B in Fig.~\ref{fig_cont}b and a finite
spherical box $0 \leq r_i \leq R$, where ${-1 \leq j \leq  N_t+1}$
  and~${0 \leq i \leq N_r-1}$. The complex field~${\phi_{j,\, i}
  \equiv \phi_{\mathrm{cl}}(t_{j},\, r_i)}$  is stored at the
lattice sites.  We considerably decrease the time steps $|t_{j+1} -
  t_j|$ near the origin~${t=0}$, i.e.\ in the   vicinity of the
  ``main'' singularity. On the other  
  hand, our spatial lattice is uniform. Practice shows that this
  choice is optimal for achieving reasonable accuracy at restricted
  computational resources.

We discretize the boundary value problem using the standard
second-order scheme, see Appendix~\ref{app:num_methods} for
details. To this end we notice that the field equation~(\ref{eq:2.4}) 
can be obtained by extremizing the classical action with the
  source term,
\begin{equation} 
  \label{eq:11}
  S_J = S[\phi] + \frac{i}{\lambda}\int d^3 \boldsymbol{x} \,
  J(\boldsymbol{x}) \, \phi(0,\, \boldsymbol{x})\,.
\end{equation}
Discretizing the latter functional, we arrive at a nonlinear
function~$S_J$ of $\phi_{j,i}$ and the lattice field equation
\begin{equation}
  \label{eq:12}
  G_{j,\, i} \equiv \frac{\partial S_{J}}{\partial \phi_{j,\, i}} = 0\,.
\end{equation}
The Dirichlet boundary condition in the infinite past~\eqref{eq:2.5}
can be imposed at the very first time site~$t = t_{-1}$~--- the point
A of the time contour:  $\phi_{-1,\,   i} = 0$. Numerical
implementation of the condition in the asymptotic 
future~\eqref{eq:2.7}  is far less trivial. In 
  Appendix~\ref{app:num_methods} we show that it relates the field
  values at the two very last time sites~${t = t_{N_t}}$
and~$t_{N_{t} +   1}$. Indeed, in the continuous
case~$a_{\boldsymbol{k}}$ and~$b_{\boldsymbol{k}}^*$ can be 
obtained by Fourier-transforming~$\phi_{\mathrm{cl}}$
and~$\partial_t\phi_{\mathrm{cl}}$ and taking appropriate linear 
combinations of their images, see Eq.~(\ref{eq:2.6}). On the
  lattice, one can express the field and its time derivative in terms
of~$\phi_{N_t,\,   i}$ and~$\phi_{N_t +   1,\,   i}$, thus turning
Eq.~(\ref{eq:2.7}) into a linear relation between them. The
last two equations for the Lagrange multipliers~$T$ and~$\theta$ are
obtained by substituting lattice versions of~$a_{\boldsymbol{k}}$ 
and~$b_{\boldsymbol{k}}^*$ into Eqs.~\eqref{eq:2.8}.
Finally, the result for $F_J$ is given by Eq.~(\ref{eq:2.9a}) 
with the discretized action.

\begin{sloppy}

To sum up, our lattice formulation of the semiclassical boundary value
problem includes ${2N_r (N_t+3) + 2}$ real nonlinear equations for
the same number of unknowns ${y_\alpha \equiv \{ \mathrm{Re}\, 
\phi_{j,\, i},\, \mathrm{Im}\, \phi_{j,\, i},\, T,\,  \theta\}}$. We
solve them using  Newton-Raphson numerical method~\cite{NR}. Namely,
suppose a crude approximation~$y^{(0)}_\alpha$ to the solution is known. Then the
correction $\delta y_\alpha = y_\alpha - y_\alpha^{(0)}$ 
satisfies the linear system
\begin{equation}
\label{eq:3.3a}
G_\alpha(y^{(0)})+\sum_{\beta}\delta y_\beta\, \frac{\partial G_\alpha}{\partial 
  y_\beta}\Big|_{y^{(0)}} = 0\,,
\end{equation}
where $G_{\alpha}$ are the left-hand sides all lattice equations:
Eq.~(\ref{eq:12}), the boundary conditions, and equations for $T$ 
and~$\theta$. Once Eqs.~(\ref{eq:3.3a}) are solved, we
refine the approximation,~${y^{(0)}_\alpha  \to y^{(0)}_\alpha +
  \delta y_\alpha}$, and 
solve them, again, until the procedure converges. Note that 
the Newton-Raphson method is very picky to the first choice of
$y^{(0)}$, but converges quadratically  fast~\cite{NR} if the
latter is sufficiently close to the solution. We will discuss
  selection of that configuration in the next section.

\end{sloppy}

Computationally, the most time-consuming part of the problem is to
solve the sparse linear system~(\ref{eq:3.3a}). We
do this using the elimination algorithm of Refs.~\cite{Bonini:1999kj,
  Demidov:2015nea} and GPU-accelerated linear algebra
package~\cite{cuda}. 

In practical computations we fixed $N_r = 256$ and varied the size
of the spatial box between $R = 100$ at $\varepsilon = 0.35$ and
$R = 6.5$ at $\varepsilon = 5$;  recall that~${m=1}$ in our
units. This allowed us to encompass long-wave parts of nonrelativistic
configurations and,  in the case of large $\varepsilon$,
resolve high-frequency modes of more compact and energetic
solutions.  The time steps  were~${|\Delta t|
    \sim 10^{-2} \div 10^{-3}}$ near the ends of the contour and
two orders of magnitude smaller near~$t=0$. We  always selected
the Minkowskian contour length~$t_{N_t+1}$ comparable to $R$
because out-waves move inside the  lightcone. The
Euclidean part $|\mathrm{Im}\, t_{-1}|$ was shorter that that by a factor of
few. The resulting temporal lattices had~${N_t =  7061\div
   12201}$ sites.

We controlled numerical precision by changing the lattice parameters
$N_r$, $R$, $N_t$, $\mathrm{Im}\, t_{-1}$,  and~$t_{N_t+1}$. This
had led to variability of the suppression exponent of order
  $10^{-3}$ in the center of the parameter region and up to 2\%
  in the worst cases. The errors were mainly coming from the finite  
size and spatial discretization  effects, while dependence on the
temporal lattice was ten times weaker. The absolute
accuracies of~$T$ and~$\theta$ were always better
  than~$10^{-2}$, and relations~(\ref{eq:2.11}) held to the
same 
precision. We monitored the conservation of energy (\ref{eq:7}) on the
Euclidean and real-time parts of the contour. It remained stable
at the 1\% level except for the cases of the lowest and highest
$\varepsilon$ when $6\div 24$\% nonconservation was observed
near~${t=0}$. Linearization of the out-waves was
checked by 
comparing the exact and linear energies, 
Eqs.~(\ref{eq:7})  and~\eqref{eq:2.8}, respectively, at~$t
  \approx  t_{N_t +1}$, i.e.\ at the point~B of the time
contour. The relative deviation of these quantities was always 
smaller than~0.4\%. 

It is worth  noting that all numerical artefacts described above
are subdominant with respect to the extrapolation errors
originating from evaluation of   the limits~${J \to +0}$, ${\lambda
n \to +\infty}$, and~$\varepsilon \to 0$. Errorbars on the
  plots display the latter inaccuracies.


\section{A way to singular solutions}
\label{sec:way_to_sing_solut}

\begin{sloppy}

The above numerical technique allows one to reconstruct the
entire branch of saddle-point configurations from a single
representative solution. Indeed, let~${y_\alpha^{(0)} =
  \{\phi_{\mathrm{cl}}^{(0)}(t,\,  \boldsymbol{x}),\, T^{(0)},\,
  \theta^{(0)}\}}$ is a solution with
parameters $\lambda n^{(0)}$, $\varepsilon^{(0)}$,
$j_0^{(0)}$, and
$\sigma^{(0)}$, where the last two values characterize the
Gaussian source~$J(\boldsymbol{x})$. Slightly 
changing one or several parameters, 
e.g.~${\varepsilon  =  \varepsilon^{(0)} + \delta   \varepsilon}$, we
numerically search for the new solution~$y_\alpha$ using
$y^{(0)}_\alpha$ as the first approximation. If the change of parameters is
small enough, the approximation is good and the iterative
method converges. After that we reload $y^{(0)}$ with the newly found
solution and repeat the procedure,  making another step in
the parameter space and finding yet another solution, etc. In this way we
can cover  all accessible parameter region with solutions. 

\end{sloppy}

The question is: where to get the very first configuration
$y^{(0)}$?  We need a physical one giving the dominant
contribution to the path integral for the probability.

It is clear that in a certain regime the quartic interactions
are irrelevant and the particles in the final state are created
by the classical source~$J(\boldsymbol{x})$. This situation is
opposite to the target limit $J\to 0$ when all particles
  are produced by the  interaction vertices and the result is
insensitive to the source profile. Ignoring the $\phi_{\mathrm{cl}}^3$
term in the field equation~\eqref{eq:2.4}, we obtain the semiclassical solution
  in the linear theory, 
\begin{equation}
  \label{eq:3.5}
  \phi_{\mathrm{cl}}^{(\mathrm{lin})}(t,\, \boldsymbol{x}) = - \int
  \frac{d^3 \boldsymbol{k}}{(2\pi)^3} \,
  \frac{J^*(\boldsymbol{k})}{2\omega_{\boldsymbol{k}}} \,
  \mathrm{e}^{i\omega_{\boldsymbol{k}} t - i \boldsymbol{k}
    \boldsymbol{x} + \theta - 2\omega_{\boldsymbol{k}} T} - i \int
  \frac{d^4 k}{(2\pi)^4} \, \frac{J(-\boldsymbol{k}) \, \mathrm{e}^{i k
      x}}{k^2 - m^2 + i0}\,,
\end{equation}
where the first term satisfies the homogeneous equation and the second
equals Feynman's Green function convoluted with the source;
$J(\boldsymbol{k})$ is a Fourier image of the latter. Hereafter we
assume that this solution is continued analytically from the
parts~${t<0}$ and~${t>0}$ of the real time axis to the
intervals A0 and 0B of the complex contour in
  Fig.~\ref{fig_cont}b, see explicit expressions 
in Appendix~\ref{sec:semicl-solut-line}. It is straightforward to check
that Eq.~(\ref{eq:3.5}) satisfies the boundary
conditions~\eqref{eq:2.5} and~(\ref{eq:2.7}) in the asymptotic past
and future. Indeed, its two terms both vanish exponentially as $t \to
+i\infty$ but represent different, positive- and
negative-frequency components of $\phi_{\mathrm{cl}}$ as $t\to
+\infty$. As usual, the parameters $T$ and $\theta$ are related  to
$\lambda n$ and~$\varepsilon$ by  Eqs.~(\ref{eq:2.8}), and the
suppression exponent $F_J$ is obtained  by
substituting Eq.~(\ref{eq:3.5}) into Eq.~(\ref{eq:2.9a}) in the
  non-interacting case; see 
Appendix~\ref{sec:semicl-solut-line} for details.

It is clear that the above linear solution is proportional to the
amplitude~$j_0$ of the source, whereas~$\lambda n$
and~$\lambda E$ are  
quadratic in the field and thus proportional to~$j_0^2$. The
interaction term~${\phi_{\mathrm{cl}}^3 \propto j_0^3}$ in the
equation is hence irrelevant in the limit of
\begin{equation}
  \label{eq:10}
  \mbox{linear theory:} \quad \qquad \mbox{$j_0 \to 0$, ~~$\lambda n
    \propto j_0^2$ ~~     at fixed $\varepsilon$ and $\sigma$\,.}
\end{equation}
We thus expect that at small $j_0$ and $\lambda n$ the
  configuration~(\ref{eq:3.5}) approximately satisfies the full nonlinear boundary value problem.   

\begin{figure}
  \centerline{\includegraphics{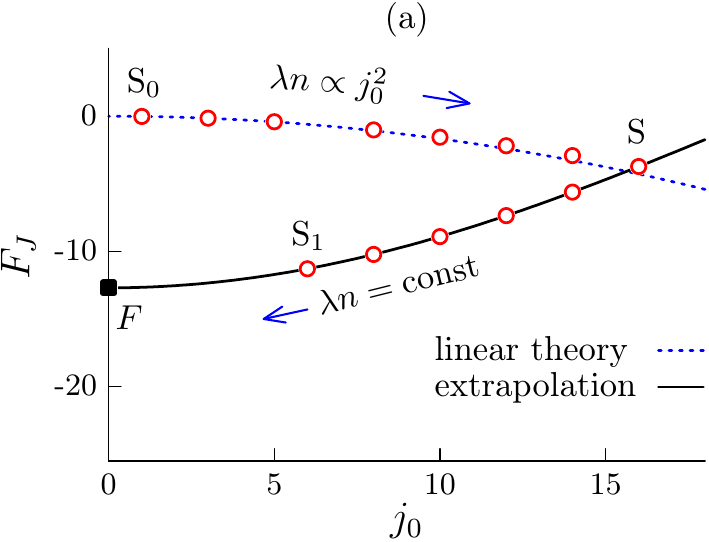}\hspace{5mm}
  \includegraphics{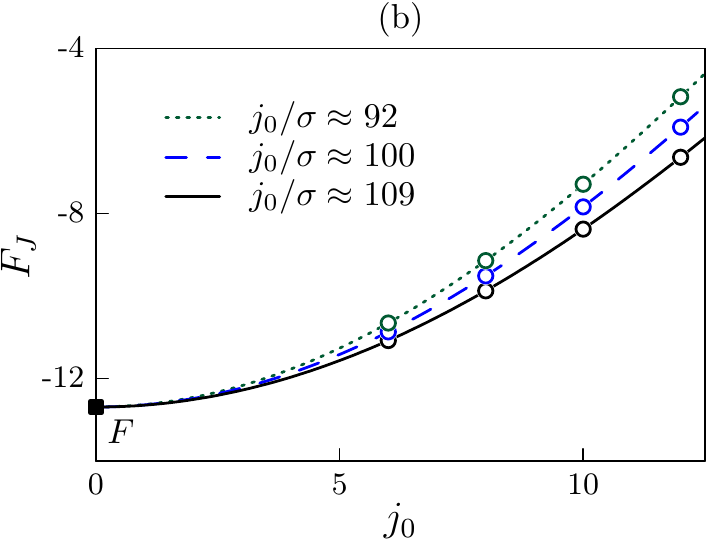}}
\caption{(a)~The exponent $F_J$ versus the source strength $j_0$ at
  $\varepsilon = 3$. The chain  S$_0$S   of
  solutions (circles) is obtained by increasing $j_0$ at given
  ${\lambda n  / j_0^2\approx 10^{-2}}$ and~$\sigma\approx
  0.13$, while the solutions from the lower branch SS$_1$
  (also circles) have fixed~${\lambda n \approx 2.51}$ and~${j_0 /
    \sigma \approx 120}$. Dotted and solid lines show the
  suppression exponent~\eqref{eq:3.8} in the linear theory and the
  polynomial extrapolation~\eqref{eq:3.11} to~${j_0= 0}$,
  respectively. (b)~Extrapolations $j_0 \to 0$ at different
  $j_0/\sigma$ for~${\varepsilon=3}$ and $\lambda n  = 2.51$ (lines
  passing through the data circles).  We use units with~${m=1}$.}
\label{fig_f_j}
\end{figure}

Importantly, the solution in the free theory is unique and definitely
physical\footnote{We checked this explicitly:
  calculated the matrix elements $\langle f; \, E,\, n | \exp\{- \int
  d^3\boldsymbol{x} \, 
    J \hat{\phi} / \lambda \}  |0\rangle$ using the algebra of
    creation and annihilation operators in the linear theory and then
    performed the final-state sum in Eq.~(\ref{eq:1.1}) via the
    steepest descent method. The resulting value of~$F_J$
    coincided with that for the semiclassical solution 
    (\ref{eq:3.5}).}, as the path integral in this case is Gaussian. We
therefore use 
  Eq.~(\ref{eq:3.5}) as the very first approximation for the
  numerical procedure described above. Namely, setting
    $y_\alpha^{(0)} = \{\phi_{\mathrm{cl}}^{\mathrm{(lin)}},\,
    T^{\mathrm{(lin)}},\, \theta^{\mathrm{(lin)}}\}$ 
    at sufficiently small $j_0$,  $\lambda n \sim O(j_0^2)$, and
  finite $\sigma$ and $\varepsilon$, we observe that the iterations
    converge fast to the nonlinear solution with the same
    parameters. Computing the suppression exponent~$F_J$ of the
    latter, we display it with the circle~S$_0$ in 
  Fig.~\ref{fig_f_j}a. After that we start increasing~$j_0$ and~$\lambda  
n\propto j_0^2$ in small steps and finding numerical
solutions at  every step, until the configurations with~$\lambda
n\sim  O(1)$ are 
obtained; see the chain of  circles S$_0$S leading to the solution S
in the figure. The configurations with large $j_0$ and $\lambda
n$ are visibly different from the ones in the free theory; we will
discuss them in the next Section. In Fig.~\ref{fig_f_j}a we
compare their suppression exponents with the prediction of the linear
theory (dotted line and Eq.~\eqref{eq:3.8} from
Appendix~\ref{sec:semicl-solut-line}). As expected, 
the two graphs are close at small $j_0$ but start to deviate at large
values of this parameter.  

Independently changing the parameters~$\lambda n$, $\varepsilon$,
$j_0$, and $\sigma$ in small steps, we reproduce the entire
continuous branch of numerical solutions and compute the
  exponent~$F_J$. 

The final but nontrivial part of our procedure consists of  evaluating
the few-particle limit~$J\to 0$ or, more precisely,
\begin{equation}
  \label{eq:13}
  \mbox{few-particle:} \quad \mbox{$j_0 \to 0$, ~~$\sigma
    \propto j_0$ ~~ at fixed $\varepsilon$ and $\lambda n$.}
\end{equation}
We have already argued that the semiclassical configurations become
singular in this limit. They cannot be resolved on the lattice. Say,
the lower branch SS$_1$ of numerical solutions in Fig.~\ref{fig_f_j}a  (circles)
is obtained from S by decreasing $j_0$ in accordance with
Eq.~(\ref{eq:13}). The last representative~S$_1$ of this branch
already has  poor precision, and solutions at even smaller~$j_0$
cannot be obtained with acceptable accuracy. We will
explicitly visualize the singularities of the solutions below. 

A way out is to extrapolate results to $j_0 = 0$ using valuable
analytic input summarized in Appendix~\ref{app:lim_j_to_0}. Indeed, it
is clear~\cite{Son:1995wz} that weak and narrow source
affects the solutions locally in the vicinity of $({t,\,
    \boldsymbol{x}) = 0}$ making them regular,
i.e.\ slightly shifting their ``main'' 
singularities~$t_*(\boldsymbol{x})$ to the lower complex time plane, see
the dashed line with the crossed circle in
Fig.~\ref{fig_cont}b. Analyzing the solutions near the singularity, we
can extract their dependence on~$j_0$.  In
  Appendix~\ref{app:lim_j_to_0} we argue that the saddle-point
  configuration itself,
its Lagrange multipliers~$T$,~$\theta$, and the exponent~$F_J$ can be
expressed as power series in $j_0^2$; e.g.,
\begin{equation}
\label{eq:3.11}
F_J(\lambda n,\, \varepsilon) = F + F_1\, j_0^2 + F_2\, j_0^4 + F_3\, j_0^6
+ \dots\,,
\end{equation}
where the coefficients $F_i(\lambda n,\, \varepsilon)$ are independent
of $j_0$.

In practice, for every chosen $\lambda n$ and $\varepsilon$ we compute
the solutions at several small values of~$j_0$ but the
same~$j_0/\sigma$. Then we fit their exponents $F_J$ and the parameters
$T$,  $\theta$ with cubic polynomials of~$j_0^2$, i.e.\ the four first 
terms in Eq.~(\ref{eq:3.11}). This procedure is illustrated in 
Fig.~\ref{fig_f_j}a where Eq.~\eqref{eq:3.11} (solid line) correctly
describes the numerical data SS$_1$ (circles). The few-particle
exponent $F$ and the respective values of $T$ and $\theta$ are given
by the first coefficients of the polynomials
(filled square $F$ in the figure).

We finish this Section with the test of the suppression exponent
  universality. Recall that the universality
conjecture~\cite{Rubakov:1992ec, Libanov:1995gh} declares
  insensitivity of~$F$ to the few-particle initial state, in
particular, to the profile of the vanishingly small
source~$J(\boldsymbol{x})$ and its relative  width~$\sigma/ j_0$. In 
Fig.~\ref{fig_f_j}b we confirm that this is the case,
indeed. Namely, performing three independent  polynomial fits
(lines) of the data with different~$j_0 / \sigma$ (circles), we
arrive at the same result for~$F$ (filled square). It is 
worth stressing that the universality assumption lies in the basis of
our semiclassical method~\cite{Son:1995wz}.

Note that the extrapolation $J \to 0$ generates the
largest errors in our numerical procedure. We estimate them by
changing the numbers of data points and~$j_0$ 
  intervals in the fits. The respective scatter of the extrapolation 
  results essentially depends on~$\varepsilon$ and $\lambda n$
and is highly sensitive to the discretization errors adding
a pseudorandom component to the data. Typically, the final result
  for $F$ is stable within~$0.7\%$ precision interval
which, however, grows to~$6\%$ at the highest and lowest~$\varepsilon$
and largest multiplicities. The accuracy of the extrapolated $T$
and~$\theta$ is better than~$8\%$ in the center of our parameter region
but deteriorates to~$13-20\%$ for the
smallest\footnote{Formally, the relative error of $T$ exceeds
  100\% at high~$\varepsilon$ where this parameter is small.}
$\varepsilon$ and~$\lambda 
n$. We display extrapolation errors with errorbars in plots
whenever they are larger than the point size.

\section{Numerical results}
\label{sec:numerical-results}

\subsection{Semiclassical solutions}
\label{sec:saddle-point-conf}
 
\begin{figure}[!thb]
  \unitlength 1.2mm

  \vspace{3mm}
  \centerline{\begin{picture}(106,31.8)(-40,11.2)
  \thicklines
  \put(-19.3,11.7){\includegraphics[width=80\unitlength]{
      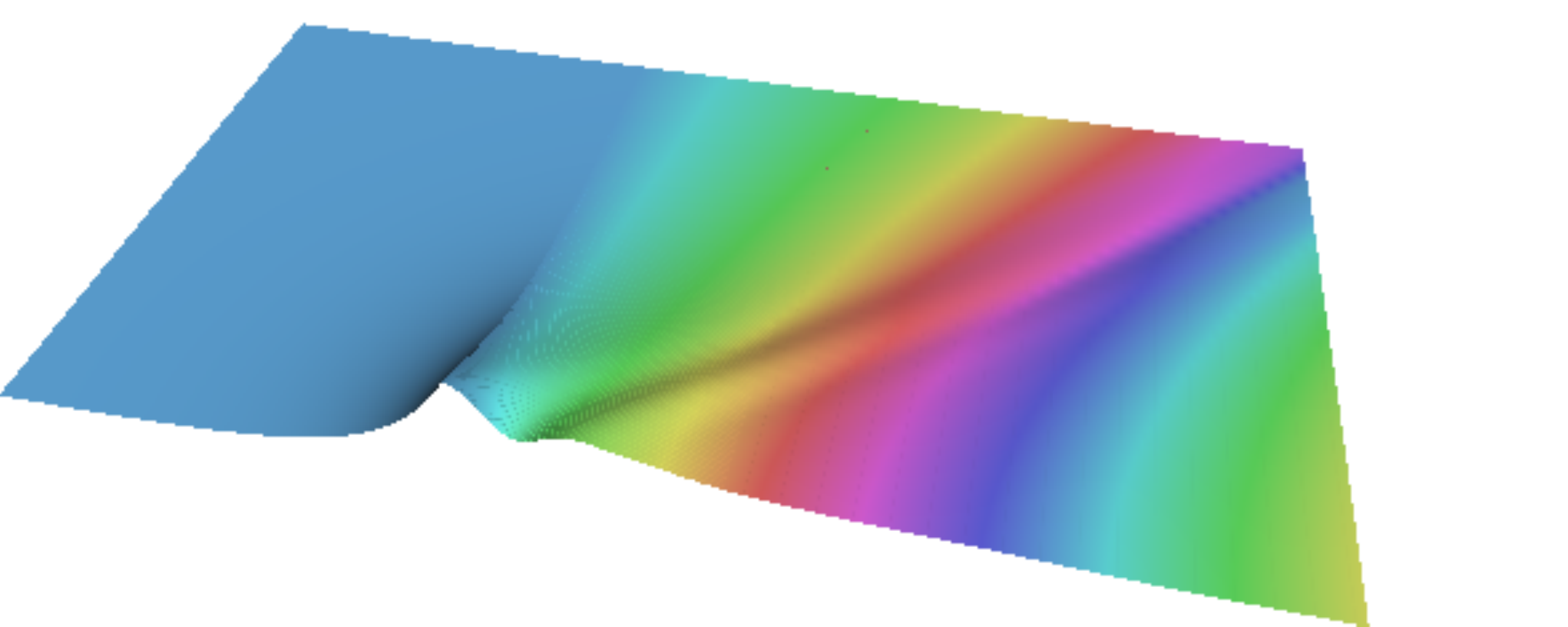}}
    \put(56.6,30){\includegraphics[width=9\unitlength]{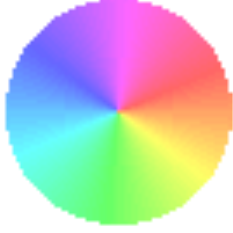}}
  \put(67,33.5){\small 0}
  \put(54.2,33.8){\small $\pi$}
  \put(57,40.5){$\mathrm{arg}\, \phi_{\mathrm{cl}}$}
  \put(-21.0,20){-4}
  \put(3.5,16.2){0}
  \put(31.0,11.5){4}
  \put(-22.0,24.0){0}
  \put(-9.5,40){5}
  \put(-32,20){\line(6,-1){20}}
  \put(-12,16.666){\vector(4,-1){1}}
  \put(-32,20){\line(4,5){15}}
  \put(-17,38.75){\vector(3,4){1}}
  \put(-32,20){\line(0,1){20}}
  \put(-32,40){\vector(0,1){1}}
  \put(-20,12.5){${\rm Re}\,t-{\rm Im}\,t$}
  \put(-20,40.5){$r$}
  \put(-40,40.5){$|\phi_{\mathrm{cl}}|$}
 \end{picture}}
 \caption{Semiclassical solution with $\lambda n\approx 0.63$,
   $\varepsilon=0.5$, $j_0=0.6$, and $\sigma\approx 0.18$ in units 
   with~$m=1$. Color indicates complex phase
   of~$\phi_{\mathrm{cl}}$.}
 \label{fig_solution_1}
\end{figure}
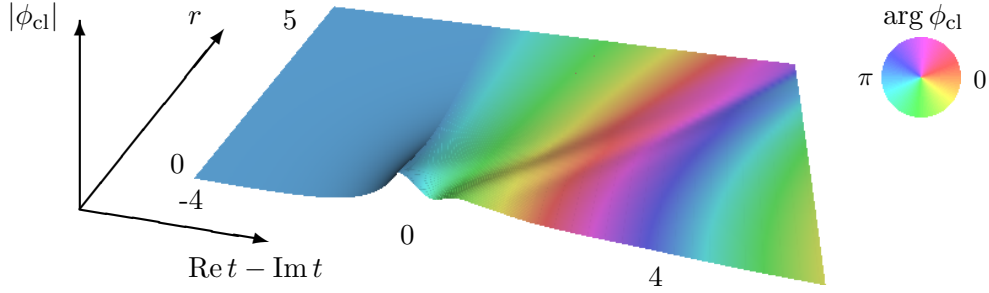

\begin{figure}
  \unitlength 0.8mm
  \begin{picture}(191,55)(-0.5,2.5)
    \thicklines
    \put(10.3,-1.4){\includegraphics[width=107\unitlength]{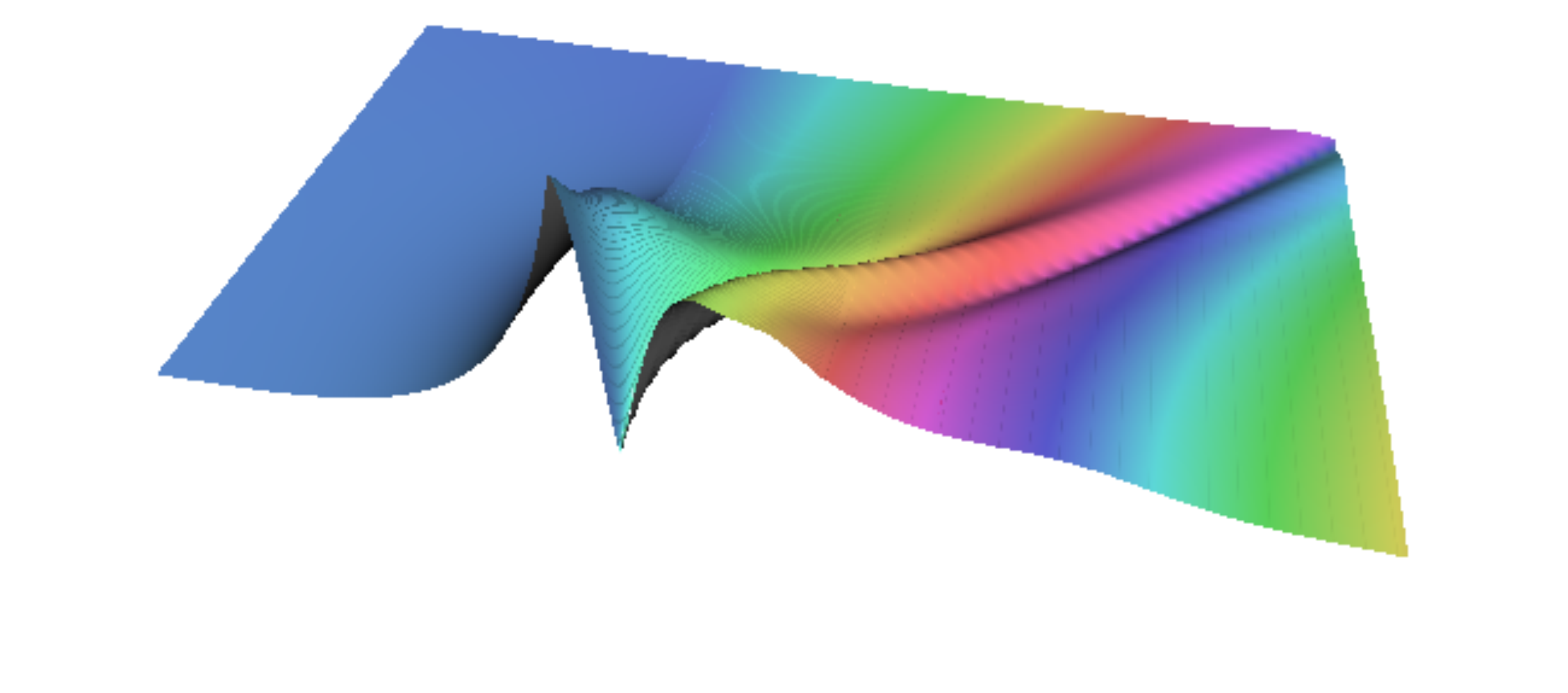}}
    \put(99.5,-3.5){\includegraphics[width=107\unitlength]{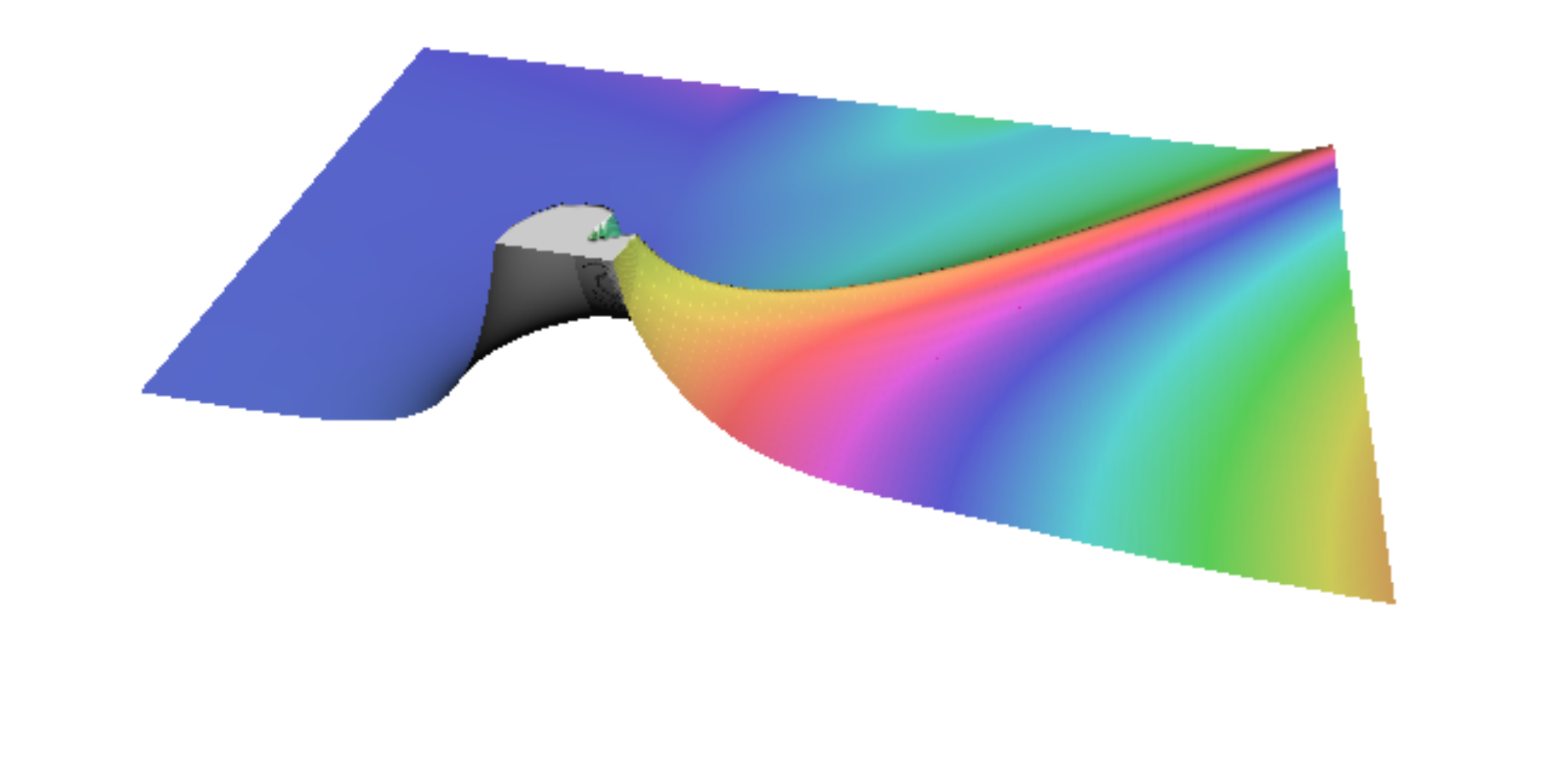}}
    \put(98.5,39.5){\includegraphics[width=12\unitlength]{pie_correct.pdf}}
    \put(112,44){\small 0}
    \put(98,54){$\mathrm{arg}\, \phi_{\mathrm{cl}}$}
    \put(95,44.5){\small $\pi$}
    \put(20,2.5){${\rm Re}\,t-{\rm Im}\,t$}
    \put(19,35){$r$}
    \put(-1,38){$|\phi_{\mathrm{cl}}|$}
    \put(18.0,15.0){-4}
    \put(47,11){0}
    \put(78.0,6.8){4}
    \put(17,20.5){0}
    \put(31.5,39){5}
    \put(106,17){-4}
    \put(138,11.5){0}
    \put(171,6){4}
    \put(104.7,22.6){0}
    \put(118,40){5}
    \put(61,48){(a)}
    \put(154,48){(b)}
    \qbezier(3,12)(13,10.571)(23,9.143)
    \put(23,9.143){\vector(4,-1){1}}
    \put(3,12){\line(4,5){15.0}}
    \put(18.5,31.0){\vector(3,4){1}}
    \put(3,12){\line(0,1){20.0}}
    \put(3,33.0){\vector(0,1){1}}    
  \end{picture}
  \caption{Two semiclassical solutions obtained from the one 
    in Fig.~\ref{fig_solution_1} by increasing (a)~the out-state
    multiplicity~$\lambda n$ and after that~--- (b)~the mean
    energy~$\varepsilon$ of out-particles. The peak of the
    solution (b) at $t,\, r  \approx 0$ is cropped off for
    visualization purposes. The parameters of the solutions are 
    (a)~$\lambda n\approx 18.8$, $\varepsilon =0.5$, $j_0=7.8$,
    $\sigma\approx 0.41$ and (b)~$\lambda n\approx
    18.8$, $\varepsilon  = 3$, $j_0=12$, $\sigma\approx 0.2$. Recall
    that the color encodes complex phase of $\phi_{\mathrm{cl}} (t,\, r)$
    and we exploit units with~${m=1}$.}
  \label{fig_solution_3}
\end{figure}

In Fig.~\ref{fig_solution_1} we display the saddle-point solution
  with relatively small out-state multiplicity ${\lambda n\approx
  0.63}$ and low kinetic energy~${\varepsilon=0.5}$, see also the  
  movie~[\citenum{movie}(a)] and recall that~${m=1}$ in   our
  units. Three-dimensional surface in this figure
shows the absolute  value of~$\phi_{\mathrm{cl}}$ as a function of
the radial coordinate~$r$ and a parameter~${\mathrm{Re}\, t
  - \mathrm{Im}\,  t}$ along the time contour\footnote{We never
show
solutions along   the deformed contour A0B$^\prime$B; cf.\ Fig.~\ref{fig_cont}b.}~A0B, while the color
  marks complex phase of~$\phi_{\mathrm{cl}}$. We see that the 
  solution decreases exponentially towards the left side of the graph,
  i.e.\ as~${t \to +i\infty}$. Besides, it has a sharp peak near the
  origin~${t = r = 0}$ where the weak source~$J(\boldsymbol{x})$
    is situated. Note that the latter source is nonzero in all our visualized
  configurations. At $t>0$
  the solution in Fig.~\ref{fig_solution_1} describes complex-valued
  outgoing wave packet that moves inside the light
  cone. Figure~\ref{fig_solution_3} demonstrates two solutions with
  other values of~$\lambda n$ and~$\varepsilon$. 

\begin{figure}[ht]
  \centerline{
    \unitlength=1mm
    \begin{picture}(72,50)
      \put(0,0){\includegraphics{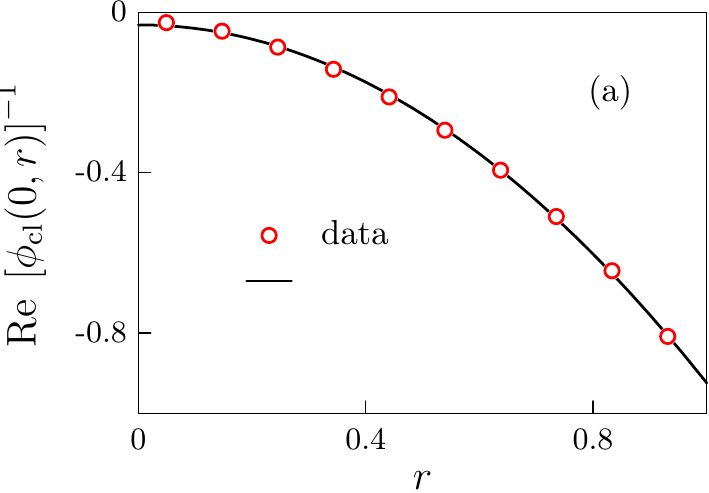}}
      \put(32.7,20.4){\small fit~\eqref{eq:24}}
    \end{picture}\hspace{5mm}
    \includegraphics{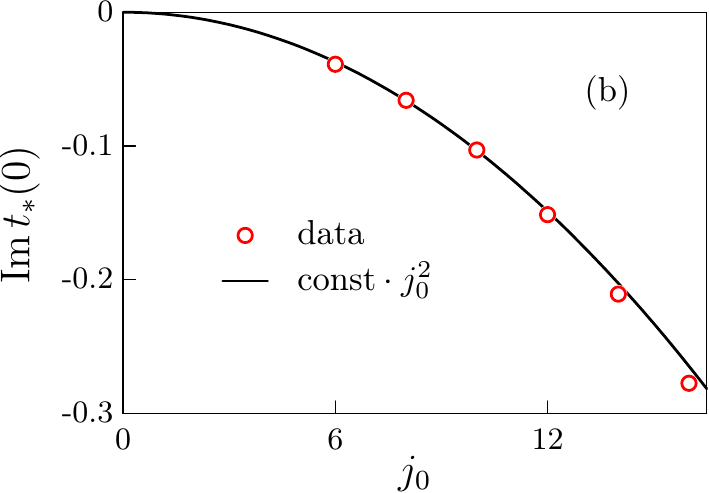}
  }
 \caption{(a) Inverse field
   $\mathrm{Re}\,\left[\phi_{\mathrm{cl}}(0,\,r)\right]^{-1}$ at $t=+0$
   and small $r$. Numerical data (circles) are fitted with 
   quadratic polynomial in Eq.~(\ref{eq:24}) (line). The solution has
   parameters~${\lambda n \approx 2.51}$,  
   ${\varepsilon = 3}$, $j_0 = 6$, and~${\sigma = 0.05}$. (b)~The
   tip~${t_*(0)}$ of the ``main'' singularity surface
   versus~$j_0$. This graph corresponds to~${j_0/\sigma =
     120}$ and the same
   out-state parameters as in Fig.~(a). The solid line is a parabola~$t_{*}(0) \propto 
   j_0^2$. Units with~$m=1$ are used in both figures.}
 \label{fig_re_phi_tau0_j}
\end{figure}

The solutions in Figs.~\ref{fig_solution_1} and~\ref{fig_solution_3}
have relatively large $j_0$ and are still far from being
singular. Their parts near the origin $t = r = 0$, however, strongly
depend on the source and turn into high and narrow
peaks once the value of $j_0$ gets smaller. Indeed, in
Appendix~\ref{app:lim_j_to_0} we derive general form of 
solutions near their singularities $t  = t_*(r)$
[cf.\ Eq.~\eqref{eq:9}], 
\begin{equation}
  \label{eq:24}
  \phi_{\mathrm{cl}} \approx \frac{-i\sqrt{2}}{t - t_*(r)}\,,
  \qquad t_*(r)  =  t_{*,\, 0} + t_{*,\, 2}\, r^2 + O(r^4) \qquad
  \mbox{at small $|t-t_*|$, $r$ and small $j_0$}\,,
\end{equation}
where  $t_{*,\, k}$ are generic Taylor coefficients. We
confirm this prediction in 
Fig.~\ref{fig_re_phi_tau0_j}a. Namely, our numerical solution
(circles) is correctly described by Eq.~(\ref{eq:24}) (line) at~${t = 
  +0}$ and small~$r$. 

Moreover, in Appendix~\ref{app:lim_j_to_0} we also argue
that ${t_{*}(0) = O(j_0^2)}$ at small $j_0$ and
finite~$j_0/\sigma$. This means that the ``tip''  of the singularity
surface marked by the cross in Fig.~\ref{fig_cont}b touches the
origin $t=0$ in the limit of vanishing source. To check the latter 
  behavior, we computed the singularities~$t_*(0)$ of
our numerical solutions. Namely, we solved the field equation along
  the imaginary time axis from~${t=+0}$ to $\mathrm{Im}\, t < 0$ until 
$\phi_{\mathrm{cl}}(t,\, 0)$ became comparatively large, and then
  fitted its time dependence with Eq.~(\ref{eq:24}). The
  resulting values of~$t_{*}(0)$
(circles in  Fig.~\ref{fig_re_phi_tau0_j}b) are proportional to
$j_0^2$ (line), indeed. 

\begin{figure}
  \centerline{\includegraphics{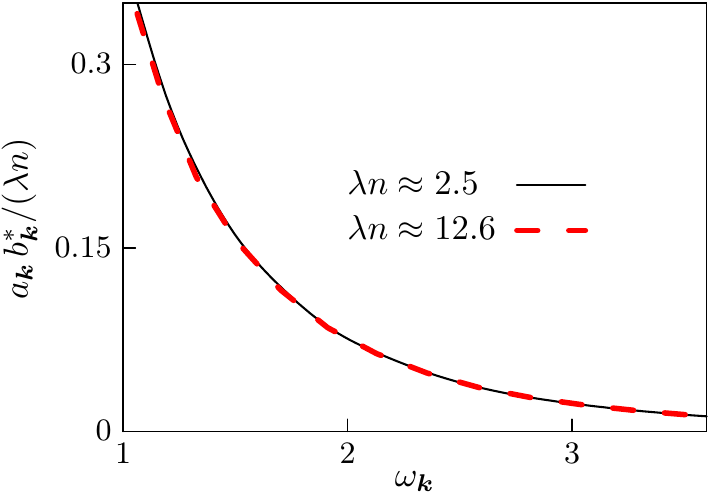}}
\caption{Rescaled occupation numbers  $a_{\boldsymbol{k}}
  b_{\boldsymbol{k}}^* / (\lambda n)$ of the final state versus
  the particle energy~$\omega_{\boldsymbol{k}}$. The two graphs  
  correspond to  $\lambda n \approx 2.51$ and $\lambda n = 12.57$ at
  $\varepsilon = 3$ and $j_0 = 0$. We use units with  $m=1$.}
  \label{fig_n_t_limit}
\end{figure}

With growth of $\lambda n$, the solutions become visibly larger
in 
size and more nonlinear in the region of finite $|t|$ and $r$, see
Fig.~\ref{fig_solution_3}a. Besides, they develop extra peaks near
the second singularity surfaces $t_*'(r)$ from the chain in
  Fig.~\ref{fig_cont}b. This last property suggests that the additional
singularities come closer to the real time axis
and start to affect the field evolution. Also, at~${\lambda n \gg 1}$ the amplitudes of
the out-waves are visibly larger. We observe the following scaling in the
linear region: 
\begin{equation}  
  \label{eq:4.7}
  \phi_{\mathrm{cl}} (t,\, r) \approx  \sqrt{\lambda n} \cdot
  \tilde{\phi}(t,\, r) \qquad \mbox{at} \qquad \mbox{$t\to +\infty$}
  \quad \mbox{and} \qquad \lambda n\gg 1\,,
\end{equation}
where $\tilde{\phi}$ does not depend on the multiplicity. Indeed,
let us have a look at the
  rescaled final-state occupation numbers~$a_{\boldsymbol{k}}
  b_{\boldsymbol{k}}^{*}/(\lambda n)$ in Fig.~\ref{fig_n_t_limit},
    where $\int d^3\boldsymbol{k} \,
    a_{\boldsymbol{k}}b_{\boldsymbol{k}}^*  = \lambda n$. Their 
  dependence on the particle  energy $\omega_{\boldsymbol{k}}$ has
  the same form at essentially 
different~$\lambda n$ in accordance with the above
prediction.

We envision that the asymptotic property~\eqref{eq:4.7} may
  be valid in a wide class of models. Besides, the distribution
  of the out-particles in  Fig.~\ref{fig_n_t_limit},
    which is independent of~$n$, may
  serve as a  
useful benchmark signature for multiparticle production, if the latter
will be ever considered in the experimental context.

\begin{figure}
  \centerline{\includegraphics{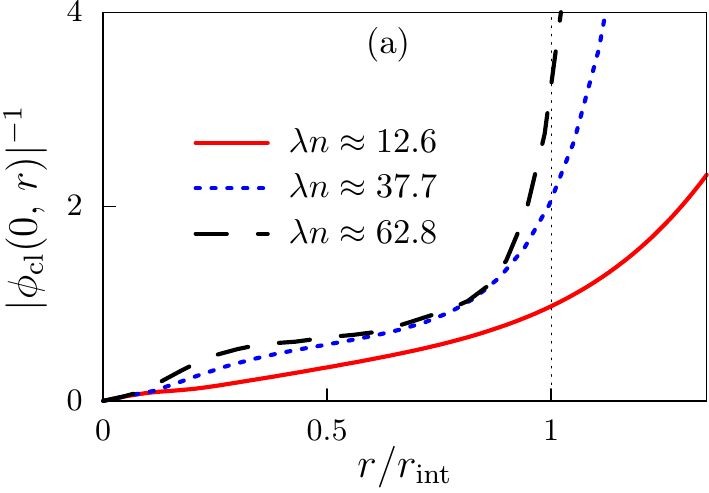}\hspace{5mm}
    \includegraphics{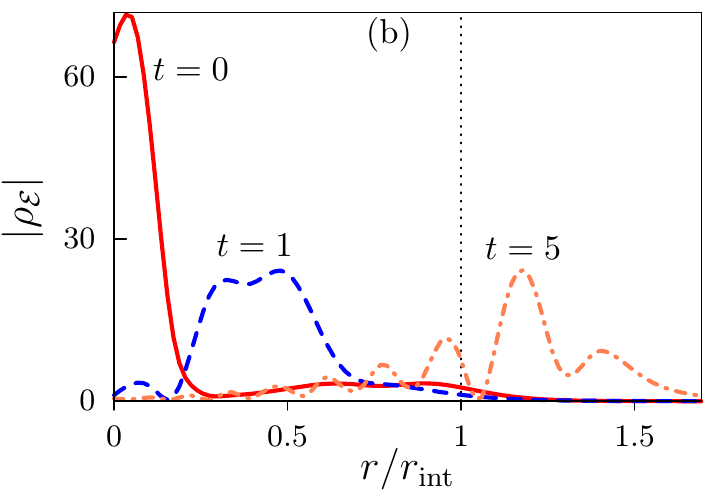}}
  \caption{(a) Inverse field $|\phi_{\mathrm{cl}}(0,\, r)|^{-1}$ as a
    function of the rescaled radial coordinate~$r/r_{\mathrm{int}}$ at
    $t=0$. All the
      graphs have~$\varepsilon=1$, $j_0 = 
      12$, and $\sigma \approx  0.19$. (b)~Energy
    density $|\rho_{\cal E}|$ depending on~$r/r_{\mathrm{int}}$
    for the solution with $\lambda n \approx 37.7$. The other 
    parameters are the same as in Fig.~(a). Units with~${m=1}$
    are used.}
  \label{fig:en_density}
\end{figure}

The above scaling is no longer valid in the interaction region where
the~$\phi^3$ term of the field equation~\eqref{eq:2.4} cannot be
neglected.  Indeed, the Ansatz~\eqref{eq:4.7} would make this 
term dominant and uncompensated in the  limit
$\lambda n \to +\infty$. We  estimate the size~$r_{\mathrm{int}}$ of
the nonlinear region by observing that at $\varepsilon
\gtrsim O(m)$ the out-waves  go away with
decreasing amplitudes~${\phi_{\mathrm{cl}}\propto \sqrt{\lambda n} \,
  /\,r}$
roughly along the lightcone~${r\sim t}$. Then the $\phi^3$ term is essential in 
\begin{equation}
  \label{eq:25}
  \mbox{the interaction region:} \quad r\lesssim  r_{\mathrm{int}} \sim
  \frac{\sqrt{\lambda n}}{(\varepsilon^2 + 2m \varepsilon)^{1/2}} \quad  \mbox{and}
  \quad  t \lesssim r_{\mathrm{int}}\,.
\end{equation}
This expectation is supported by our
numerical results. In particular, Fig.~\ref{fig:en_density}a
shows the inverse field $|\phi_{\mathrm{cl}}(0,\, r)|^{-1}$ at
$t=0$  as a function of $r/r_{\mathrm{int}}$ for three  large values
of~$\lambda n$. With growth of multiplicity, the graphs at
$r<r_{\mathrm{int}}$ approach a particular almost flat profile
of
  height~${\phi^{-1}_{\mathrm{cl}} \lesssim m^{-1}}$. Recall that
 small value of $\phi_{\mathrm{cl}}^{-1}(0,r)$ estimates the
position~$t_*(r)$ of the singularity surface which is,  therefore, also
flat at large $\lambda n$ and~${r \lesssim    O(\lambda n)^{1/2}}$; see
Eq.~(\ref{eq:24}). At $r\sim r_{\mathrm{int}}$  and beyond this point,
the graphs in Fig.~\ref{fig:en_density}a sharply increase indicating
entrance into the linear region with small~$\phi_{\mathrm{cl}}$. 

We visualize the nonlinear stage of evolution by showing the energy   
density~$|\rho_{\cal  E}|$ of the solution~--- the integrand in
Eq.~(\ref{eq:7})~---   at different moments of
time~$t$ in
Fig.~\ref{fig:en_density}b. Apparently,
the source and the nearby singularity create a huge localized
peak of $\rho_{\cal E}$ at~${t=0}$ which evolves to larger~$r$
remaining narrow and essentially interacting until it crosses the 
boundary~${r_{\mathrm{int}}\propto \sqrt{\lambda n}}$ of the
  nonlinear region (dotted vertical line). At~${r \sim 
r_{\mathrm{int}}}$ the peak quickly linearizes and starts to satisfy
Eq.~\eqref{eq:4.7}, see the graph with~${t=5 m^{-1}}$ in
Fig.~\ref{fig:en_density}b. Such nonlinear evolution of a narrow shock
supports the ``thin wall'' Ansatz suggested in
  Ref.~\cite{Gorsky:1993ix} and used in the papers on
``Higgsplosion''~\cite{Khoze:2017ifq, Khoze:2018mey}. 
 
Figure~\ref{fig_solution_3}b demonstrates the solution
with high mean energy~$\varepsilon$ of the out-particles. This
configuration is essentially sharper and has narrow outgoing wave
packet localized on the light cone~${r\sim t}$. One may assume that
such solutions with $\varepsilon \gg m$ can be obtained in the
massless theory, and the parameter~$m^2$ can only cause their small
deformation. We will further justify this observation in the next
Section.  


\subsection{Suppression exponent}
\label{sec:numer_results}

\begin{figure}
  \centerline{
    \begin{minipage}{7.2cm}
      \unitlength=1mm
      \begin{picture}(72,52)
        \put(0,0){\includegraphics{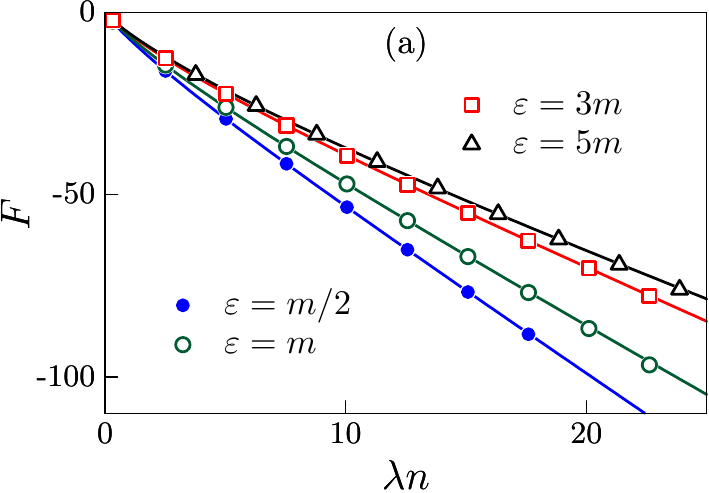}}
      \end{picture}
    \end{minipage}\hspace{7mm}
    \begin{minipage}{7.2cm}
      
      \vspace{3.5mm}
      \centerline{(b)}
      
      \vspace{3mm}
      \centerline{\begin{tabular}{cccc}
        \hline
        $\varepsilon$ & $f(\varepsilon)$ & $f_{\infty}(\varepsilon)$ &
        $g_{\infty}(\varepsilon)$\\
        \hline
        $m/2$ & $-3.46$ & $-4.47 \pm 0.20$ & $-9.7 \pm 3.8$\\
        $m$    & $-2.81$ & $-3.66 \pm 0.31$ & $-17.0 \pm 10.4$\\
        $3m$  & $-2.11$ & $-2.88 \pm 0.19$ & $-13.8 \pm 5.3$\\
        $5m$ &  $-1.92$ & $-2.35 \pm 0.54$ & $-32.8 \pm 29.2$
      \end{tabular} }
      \vspace{17mm}
  \end{minipage} }
  \caption{(a) Suppression exponent $F$ as a function of $\lambda n$
    at  fixed~$\varepsilon$. Points demonstrate numerical data, lines are
    the fits~\eqref{eq:4.10s}. (b)~Parameters in Eq.~\eqref{eq:4.10s}.}
\label{fig_f_n_log_and_two_regimes}
\end{figure}

We had already presented main results for the exponent $F(\lambda n,\,
\varepsilon)$ in the Introduction. Here we study its asymptotic, fit
the data with convenient formulae, and show remaining numerical
results in Fig.~\ref{fig_f_n_log_and_two_regimes}a (points). Recall
that we confirmed universality of the exponent in
Sec.~\ref{sec:way_to_sing_solut} and Fig.~\ref{fig_f_j}b. 

Start with the regime $\lambda n\ll 1$. It corresponds to 
perturbative limit because the series in~$\lambda n$ can be viewed as
expansion in~$\lambda$. Main contributions in this case come
from the tree-level diagrams giving~\cite{Brown:1992ay, Libanov:1994ug, 
  Son:1995wz},
\begin{equation} 
\label{eq:4.1}
\mbox{tree:} \qquad F(\varepsilon, \lambda n) = \lambda n \ln{\frac{\lambda n}{16}}-\lambda
n + f(\varepsilon)\, \lambda n +O\left(\lambda n\right)^2\,,
\end{equation}
where $O(\lambda n)^2$ contains loop corrections
and the function $f(\varepsilon)$ is computed numerically
in Ref.~\cite{Bezrukov:1998mei}. We determine $f(\varepsilon)$ using 
half-analytic~O(4) approximation of Ref.~\cite{Bezrukov:1995ta}
which
works extremely well~\cite{Bezrukov:1998mei} in our parametric
region~$\varepsilon \leq  5m$. Results for this function are
tabulated in the first  column of
Fig.~\ref{fig_f_n_log_and_two_regimes}b. The overall tree-level
exponent~(\ref{eq:4.1}) is shown at $\varepsilon=3m$  in
Fig.~\ref{fig:intro_results}a  (dotted
line). As  expected, it is close to the  numerical data (circles) at~$\lambda n \ll 1$. 

We performed more indicative comparison with the perturbative results
in the accompanying paper~\cite{Demidov:2021rjp}. In there, we extracted
$f(\varepsilon)$ by fitting the numerical data for~$F$  at
small~$\lambda n$ with Eq.~(\ref{eq:4.1}). The result  agreed with
the tree-level exponent of Refs.~\cite{Bezrukov:1995ta, Bezrukov:1998mei}.

In the opposite limit of large $\lambda n$ the tree-level
exponent~(\ref{eq:4.1}) becomes positive which may be naively 
taken~\cite{Ringwald:1989ee, Espinosa:1989qn} for a signal  of
unsuppressed multiparticle production. But in fact, the value of
  $F$ is dominated at $\lambda n \gtrsim 1$ by loop corrections and has the  
opposite behavior: it decreases monotonically
with the multiplicity and approaches the linear asymptotic~(\ref{eq:4.9})
at~${\lambda n \gg 1}$, see Fig~\ref{fig:intro_results}a. The
slope~$f_\infty(\varepsilon) < 0$ and  shift~$g_{\infty}(\varepsilon)$
of the asymptotic strongly 
depend on $\varepsilon$, cf.\  Fig.~\ref{fig:intro_results}b. In
practice, it is convenient to approximate the numerical  data
at finite $\lambda n$ and $\varepsilon$ with the interpolating
formula 
\begin{equation}
\label{eq:4.10s}
F \approx \lambda n f_{\infty}(\varepsilon) - \frac{\lambda n}{2} \, \ln\left[
  \left(\frac{16}{\lambda n}\right)^2 \mathrm{e}^{2 - 2f (\varepsilon) + 2f_\infty(\varepsilon)} -
  \frac{2g_{\infty}(\varepsilon)}{\lambda n} + 1 \right]\,,
\end{equation}
which reduces to  Eqs.~(\ref{eq:4.1}) and (\ref{eq:4.9}) in the
limits of small and large multiplicity.  Indeed, fits with
Eq.~(\ref{eq:4.10s}) (lines in
Fig.~\ref{fig_f_n_log_and_two_regimes}a) pass through all the data
points. The best-fit values of~$f_{\infty}(\varepsilon)$
and~$g_{\infty} (\varepsilon)$ are tabulated in 
Fig.~\ref{fig_f_n_log_and_two_regimes}b and plotted  
in Fig.~\ref{fig:intro_results}b. We checked that they  are
consistent with the results of simple linear fits\footnote{We
  performed another strong test of
Eq.~(\ref{eq:4.10s}). Relations~\eqref{eq:2.9a}
  and~(\ref{eq:2.11}) give the saddle-point value of the
classical action in terms of the exponent and its $\lambda
n$ derivative at   $\varepsilon = \mbox{const}$:
${2\lambda\mathrm{Im}\, S[\phi_{\mathrm{cl}}] = F - 
  \partial F / \partial \ln (\lambda  n)}$,  where  $J=0$. This
  expression turns 
  Eq.~(\ref{eq:4.10s}) into an interpolating formula for~$2\lambda
  \mathrm{Im}\, S$. Approximating the numerical data at
  $J=0$ with the latter, we extracted 
  $g_{\infty}(\varepsilon)$ and~$f_{\infty}(\varepsilon)$ which agreed
  with the values in
  Fig.~\ref{fig_f_n_log_and_two_regimes}b.}~\eqref{eq:4.9} at 
$\lambda n \gg 1$. Note also that~$f(\varepsilon)$ is fixed by an
  independent tree-level calculation and remains constant in the
  fits.

Now, consider the limit $\varepsilon \to 0$ in which all final
particles are produced at the mass threshold. The scattering
amplitude~${\cal
  A}_n$ is expected to have a finite limiting value
  corresponding to zero outer momenta~\cite{Brown:1992ay,
  Argyres:1992np,  Smith:1992kz,   Libanov:1993qf,
  Khoze:2014zha}. Then the inclusive probability 
factorizes at low~$\varepsilon$ into~${{\cal P}_n \approx |{\cal
    A}_n|^2\, {\cal V}_n/n!}$, where~\cite{Son:1995wz}
\begin{equation}
\label{eq:4.6a}
  \frac{{\cal V}_n}{n!}\approx \frac{m^{2n-4}}{n!} \,
  \exp\left\{\frac{3n}{2}\, \ln\left(\frac{\varepsilon}{3\pi m} \right) +
  \frac{3 n}{2} - n \ln 2 + \frac{n\varepsilon}{4m}\right\}\,, \qquad \qquad \varepsilon \ll m
\end{equation}
is the total phase volume of $n$ identical nonrelativistic
particles. Using the above observation, we extract the
exponent $F_{\cal A}$, Eq.~\eqref{eq:4}, of the amplitude from the
probability as follows:
\begin{equation}
  \label{eq:26}
  F_{\cal A}(\lambda n ) = \frac12 \lim_{\varepsilon \to +0} \left[\,
    F(\lambda n,\, \varepsilon)
  - \lambda \ln ({\cal V}_n m^{4-2n}) \, \right]\,.
\end{equation}
With the ideal data, one might be able to evaluate this limit
directly, by fitting the combination in  the right-hand side with
polynomials of $\varepsilon$ and 
extracting constant terms. 
But that is hard to do in practice, since our values of $F$ already
have essential inaccuracies due to previous extrapolation~$ j_0 
\to 0$. We increase precision by recalling that the semiclassical
procedure conveniently provides the $\varepsilon$ derivative
of the exponent~${\partial F / \partial \varepsilon = 2\lambda n T}$,
see Eqs.~\eqref{eq:2.11}. Recall that we
obtain the values of $T$ 
on par with the numerical solutions, and we also extrapolate them 
to~$j_0=0$. It is straightforward to see that
the~$O(\varepsilon)$ term cancels out in the Taylor
series expansion of the combination 
\begin{equation}
\label{eq:d.7}
F  - \lambda \ln ({\cal V}_nm^{4-2n}) - 2\lambda n \varepsilon T+
\lambda n ( 3/2  + \varepsilon/ 4m)
= 2F_{\cal A} + \varepsilon^2 F_{{\cal A},\, 2} + \varepsilon^3
F_{{\cal A},\, 3} + O(\varepsilon^4)
\end{equation}
because ${
  \lambda \varepsilon\partial_\varepsilon \ln  {\cal V}_n \approx
  \lambda n (3/2 + \varepsilon/4m)}$ according
to Eq.~(\ref{eq:4.6a}). Note that $F_{\cal A}$ in the right-hand side of
Eq.~(\ref{eq:d.7}) is our target exponent of the threshold
  amplitude and we denoted the other expansion coefficients by  
$F_{{\cal  A},\, i}$. In practice, the three-parametric fit of 
  the quantity in the left-hand side of Eq.~(\ref{eq:d.7}) with the
  sparse polynomial in the  right-hand side is much 
more stable and leads to smaller errors than direct numerical
  evaluation of Eq.~\eqref{eq:26}. 

We thus arrive at the amplitude exponent~$F_{\cal A}(\lambda n)$ 
shown by the circles in Fig.~\ref{fig_cont}a. At~${\lambda n \lesssim 10}$
these data are close to the perturbative expression (dashed line),
\begin{equation}
\label{eq:4.4}
F_{\cal A} = \frac{\lambda n }{2} \big[\ln (\lambda n/8) - 1\big]
  - \frac{ (\lambda n)^2 \,3^{3/2}}{32\pi^2} \,\ln (2 +
  \sqrt{3})   + O(\lambda n)^3 \qquad \mbox{at} \qquad \varepsilon=0\,,
\end{equation}
which includes the tree-level result~\cite{Brown:1992ay} (dotted line)
and one-loop correction~\cite{Voloshin:1992nu, Libanov:1997nt} in
  the first and second terms, respectively. In the opposite 
case of large $\lambda n$ we  expect linear asymptotic ${F_{\cal A}
  \to  \lambda n  f_{\infty}'    + g_{\infty}'}$ previewed in
Eq.~(\ref{eq:5}). Hence, it is convenient to merge 
small- and large-$\lambda n$ behavior in a single interpolating
formula [cf.\ Eq.~\eqref{eq:4.10s}]
\begin{equation}
\label{eq:4.5a}
 F_{\cal A} = \lambda n f_\infty' - \frac{\lambda n}{4} \ln \left[
 \left(\frac{8}{\lambda n}\right)^2\mathrm{e}^{2+4f_{\infty}'} - 
 \frac{4 g_\infty' }{\lambda n} + 1\right]\,,
\end{equation}
which correctly describes all the numerical results (solid line in
Fig.~\ref{fig_cont}a).  Best-fit values of~$f_{\infty}'$
and~$g_{\infty}'$ are given in Eq.~(\ref{eq:5}). 

\begin{figure}
  \centerline{\includegraphics{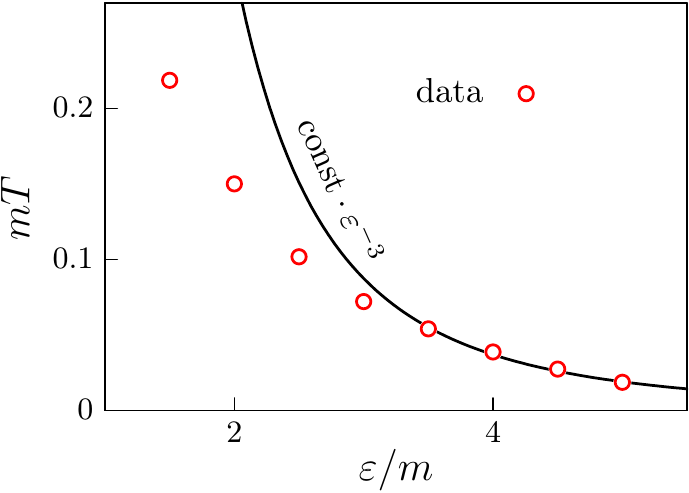}}
  \caption{Lagrange multiplier~$T$ as a
  function of $\varepsilon$ at  $\lambda n \approx 18.8$ and $j_0 = 0$
  (circles). Solid line demonstrates the asymptotic $T  \propto
  \varepsilon^{-3}$  at large $\varepsilon$.}
  \label{fig:Teps}
\end{figure}

Now, consider the limit of highly ultrarelativistic particles in the
final state $\varepsilon \to +\infty$. The respective numerical solutions
 are sharper, and their time and space derivatives
visibly grow with~$\varepsilon$, see
Fig.~\ref{fig_solution_3}b. In this regime it is
natural to treat the parameter $m^2$ in the
field equation perturbatively. On dimensional grounds one finds,
\begin{equation} 
  \label{eq:27}
  T = \frac{T_{-1}}{\varepsilon} + T_3 \, \frac{m^2}{\varepsilon^3} +
  O(\varepsilon^{-5})\,, \quad  \mbox{hence} \quad F = F_{0} +
  2\lambda n T_{-1} \, \ln\frac{\varepsilon}{m} - \lambda n T_{3} \,
  \frac{m^2}{\varepsilon^2} + O(\varepsilon^{-4})\,,
\end{equation}
where the dimensionless coefficients $T_i$ in the series depend on
$\lambda n$ and we used the Legendre relation ${\partial F / \partial
  \varepsilon = 2\lambda n T}$ in the second equality, cf.\ 
Eqs.~(\ref{eq:2.11}). As we will prove later, $F$ cannot decrease
with energy, i.e.\ $T\geq 0$ and $T_{-1} \geq 0$. This is compatible
with Eq.~(\ref{eq:27}) only if~$T_{-1}=0$: otherwise, the
suppression exponent would be positive and break unitarity at
sufficiently high energies. We  conclude that at
$\varepsilon \to +\infty$ the exponent is
$\varepsilon$-independent  and~${T \propto \varepsilon^{-3}}$. The
latter scaling  is confirmed in Fig.~\ref{fig:Teps}. 

Since parametric behavior of the exponent at low and
high~$\varepsilon$ is fixed, we can construct an interpolating
formula for its dependence on energy or, equivalently, for 
  its tilt~${f_{\infty} = f_{\infty}(\varepsilon)}$  at large
$\lambda n$. In 
practice we fit $f_{\infty}(\varepsilon)$ with the expression~\eqref{eq:4.15}
that approaches a constant as
$\varepsilon \to +\infty$ and a logarithm $\frac{3}{2} \ln\,
(\varepsilon /m)$ from the phase volume at low $\varepsilon$. With the
appropriately chosen parameters, this formula correctly  describes all the
numerical data, cf.\ the circles and the
solid line in Fig.~\ref{fig:intro_results}b.

We finish this Section with a remark that our numerical data are not
limited to the four lines in Fig.~\ref{fig_f_n_log_and_two_regimes}a. In
the ancillary files~\cite{ancillary} we provide extra raw data at
$\varepsilon/m = \{0.35,\, 0.75,\, 1.5,$ $2,$ $2.5,$ $3.5,$ $4,$ 
$4.5\}$ and different $\lambda n$, $j_0$, $\sigma$, results of their
extrapolation to $j_0=0$, values of~$f(\varepsilon)$ and best-fit
results for $f_{\infty}$ and~$g_{\infty}$. 

\section{Discussion}
\label{sec:conclus}
In this paper we computed the probabilities of 
processes $\mbox{few} 
\to n$ in the unbroken $(3+1)$-dimensional $\lambda\phi^4$ theory. To
this end we numerically implemented D.T.~Son's semiclassical method of
singular solutions. Our data cover a wide  range  of final-state
multiplicities~${n\gg 1}$ and  total collision energies~$E$. They
show that the multiparticle probabilities  fall off
monotonically with~$n$ and approach the decreasing exponent
(\ref{eq:4.9}) at~${n \gg \lambda^{-1}}$. Up to our knowledge, no
consistent calculation of this kind was performed before in a
full-fledged field-theoretical model. 

We have already presented main results in the Introduction. Here,
we critically analyze their consistency. First, the
probability~(\ref{eq:1.1}) cannot exceed unity:
\begin{equation}
  \label{eq:28}
  {\cal P}_n 
  \sim \mathrm{e^{F/\lambda}}\leq 1\,, \qquad
  \mbox{and hence} \qquad F(\lambda n,\, \varepsilon) \leq 0\,.
\end{equation}
 Note that breaking of
Eq.~(\ref{eq:28}) in any parametric region would undermine 
credibility of the entire method; cf.\ Eq.~(\ref{eq:1}). But in
reality it is satisfied by all our numerical data. In 
particular, the asymptotic of the exponent at  large~$\lambda n$ is
negative: $F \to \lambda n \, f_{\infty}(\varepsilon) +
g_\infty(\varepsilon) < 0$ at $\lambda n  \to +\infty$, see
Fig.~\ref{fig:intro_results}b and table in
Fig.~\ref{fig_f_n_log_and_two_regimes}b.

It is worth reminding that our semiclassical method relies on  the
universality conjecture~\cite{Libanov:1994ug, Libanov:1995gh,
  Jaeckel:2018ipq, Jaeckel:2018tdj, Rubakov:1992ec, Tinyakov:1991fn,
  Mueller:1992sc, Bonini:1999kj, Levkov:2008csa} for the exponent in
Eq.~(\ref{eq:28}). Namely,  the value of $F(\lambda n,\, \varepsilon)$
does not depend on the details of the initial state as long as the 
latter includes few, i.e.\ $\ll \lambda^{-1}$, particles. We
explicitly tested this assumption in Sec.~\ref{sec:way_to_sing_solut},
see Fig.~\ref{fig_f_j}b.  Its consequence is that the cross
section~${\sigma_n \propto   \exp(F/\lambda)}$ of~$2\to n$ scattering
is suppressed by the same universal function~$F$ as the
probability~${\cal P}_n$. Indeed, consider a collision of two
particles in a particular state described by wave packets with large
spatial extent $L \gg n/E$. This collision creates $n$ quanta
with the probability\footnote{This argument is rougher than the famous
Froissart bound~\cite{Froissart:1961ux}, as it assumes that the
transition amplitude is insensitive to the scattering momenta at
scales below~$L^{-1}$. But it applies to our processes at a given~$n$ and
  large enough~$L$ because the inequality $|L^{-1}\,\partial_E \ln{\cal 
  A}_n| \lesssim 1$ leads to~$L E \gtrsim |d\ln {\cal A}_n / d\ln
E|  \sim (E \partial_E F)/\lambda \sim O(n)$.}~${\cal
  P}_n \sim \sigma_n/(\pi L^2)$, i.e.\ the same exponential
  suppression. The inequality~(\ref{eq:28}) then means  that the
physical cross section $\sigma_n$ cannot be exponentially 
large. 

Second, consider the exotic process of two independent few-particle
collisions creating~$n_1$ and~$n_2$ particles in two spatially separated
regions. The overall probability for this to happen is~${\cal
  P}_{n_1}(E_1) {\cal  P}_{n_2}(E_2)$, where $E_1$ and~$E_2$  are the
respective energies. In fact, such
a two-collision event
can be regarded as a subprocess  contributing to the inclusive
probability~(\ref{eq:1.1}). Indeed, its initial state is not important
by the universality conjecture and the final state including two widely
separated particle sets is  exclusive. Since ${\cal P}_n(E)$ is larger
than the probability of a subprocess, we conclude~\cite{Demidov:2018czx},
\begin{equation}
  \label{eq:30}
   F(\lambda n_1 + \lambda n_2,\,
  E_1 + E_2) \geq F(\lambda n_1,\, E_1)  + F(\lambda n_2,\, E_2)\,,
\end{equation}
where the exponents are now expressed as
  functions of~$E$ instead of $\varepsilon$. 

\begin{sloppy}

Using Eq.~\eqref{eq:30}, it is easy to show that $F$ grows with
energy. Indeed, take~${\lambda n_2\ll 1}$. Then the second
collision is not exponentially suppressed at any~$E_2$:~$F(\lambda n_2 ,\, E_{2}) \sim  O(\lambda
n_2)$ according to Eq.~\eqref{eq:4.1}. The
inequality~\eqref{eq:30} transforms into ${F(\lambda 
  n_1,E_1+E_2)\geq F(\lambda  n_1,E_1)}$ implying that~${\partial_{E} F
\propto T}$ is positive. Our numerical results  do 
satisfy this criterion. Specifically,  $f_\infty(\varepsilon)$ in Fig.~\ref{fig:intro_results}b
grows with  $\varepsilon$ and approaches the maximal 
value~$f_\infty \to -2.57$ at~$\varepsilon \to  +\infty$.

\end{sloppy}

Another particular case of Eq.~(\ref{eq:30}) corresponds to a fixed mean
energy of final particles~$E_1 / n_1 = E_2 / n_2 =  \varepsilon +m$
at arbitrary multiplicities. We obtain the inequality
\begin{equation}
  \label{eq:32}
  F(\lambda n_1 + \lambda n_2 ,\, \varepsilon) \geq F(\lambda n_1,\,
  \varepsilon) + F(\lambda n_2,\,
  \varepsilon)\,,
\end{equation}
which means that the negative exponent cannot decrease at
  large multiplicities faster than linearly. Indeed, the
power-law behavior~${F\propto  - (\lambda   n)^{\gamma}}$ is
consistent with Eq.~(\ref{eq:32}) at~${\lambda n   \to +\infty}$ only
if~${\gamma \leq 1}$. Our numerical calculation strongly suggests
linear asymptotic ${F\to \lambda n f_{\infty} + g_\infty}$  that saturates this last
condition. Then Eq.~(\ref{eq:32}) 
reduces  to~${g_{\infty}(\varepsilon) \leq 0}$ which is also true for
our data, see the table in Fig.~\ref{fig_f_n_log_and_two_regimes}b. 

\begin{figure}
  \centerline{\includegraphics{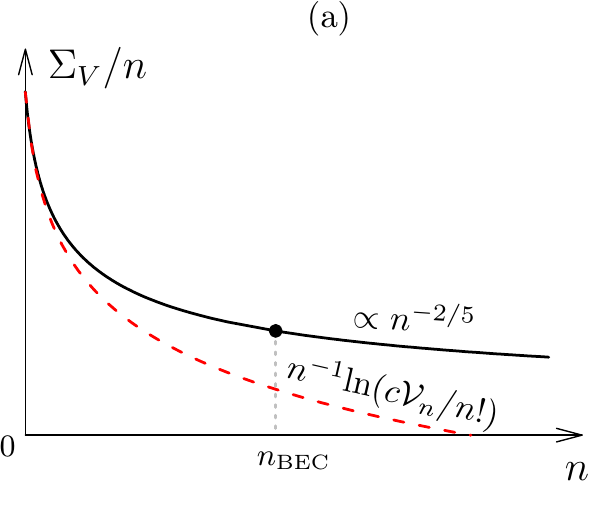}\hspace{0mm}
    \includegraphics{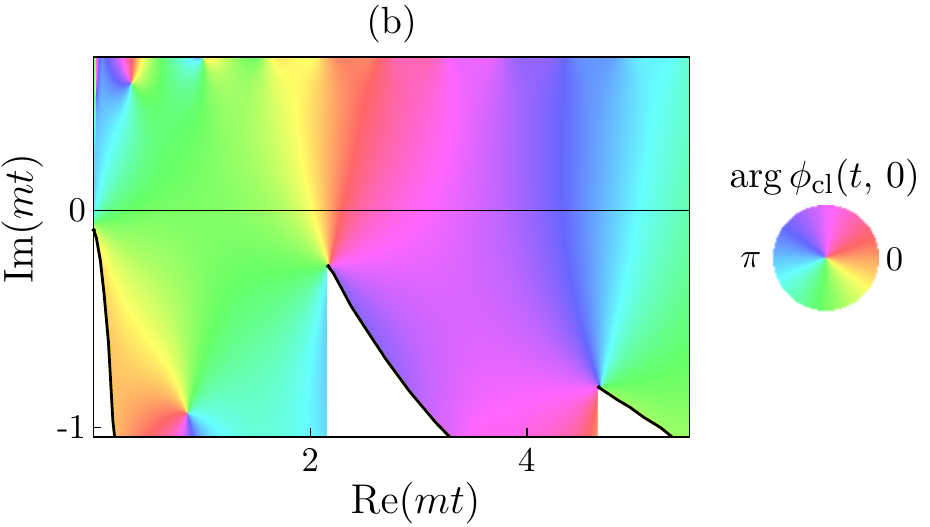}}
  \caption{(a) Entropy $\Sigma_V(n,\, E)$ of $n$ free nonrelativistic
    particles with total energy~${E \approx nm}$ in a finite
    volume~$V$ (solid line, not to scale). Dashed line is the 
    logarithm of the phase volume~\eqref{eq:4.6a}. (b)~Complex
    phase $\arg  \phi_{\mathrm{cl}}(t,\,  0)$ of the saddle-point
    configuration as a function of complex time at~${r=0}$
    (color). Black solid lines indicate singularities
    of~$\phi_{\mathrm{cl}}$. We do not perform computations in the
    white regions below them. The solution has
    parameters~${\lambda n =
      2.51}$, ${\varepsilon =  m/2}$, ${j_0 = 0.053 \, m^2}$, and~${\sigma =
      0.29 \, m^{-1}}$.}
  \label{fig_entropy}
\end{figure}

Third, one may be surprised by the fact that the
amplitude~(\ref{eq:5}) of creating $n$ particles at the mass threshold
still grows factorially with~$n$ at~$\lambda n \gtrsim 1$.  This
effect is purely kinematical and consistent with unitarity: recall that
the amplitude was extracted from the exponentially small
probability. Indeed, the factor $n!$ comes from the phase volume~${\cal
  V}_n/n!$ which has questionable physical interpretation in
the limit~${n \to  +\infty}$. To see this, consider a finite
spatial volume~$V$.  The number of nonrelativistic $n$-particle states 
in that region is given by the exponent of the thermodynamical
entropy~$\exp\{\Sigma_V(n,\, E)\}$ (solid line in
Fig.~\ref{fig_entropy}a). The latter, in turn, is
proportional to the phase 
volume $\exp(\Sigma_V)  = c{\cal V}_n/ n!$ with 
coefficient ${c \sim (2mV)^{n}/m^4}$ (dashed line in the figure), but
only at low multiplicities. At  large $n \gtrsim n_{\mathrm{BEC}}$ the wave
functions of the gas particles start to overlap, Bose-Einstein
condensation occurs, and the entropy stops being related to~${\cal
  V}_n/n!$ at all. To the contrary, it grows slowly~\cite{Landau:1980mil},
as~${\Sigma_V}   \propto n^{3/5}$, with~$n$  because new states
reluctantly appear in the overpacked Bose gas. In such a situation,
the probability~${\cal P}_n \,\mathrm{exp}(-\Sigma_V)$ of transition
to a given finite-volume state   has almost the same  suppression~$F 
\sim  \lambda n f_{\infty} < 0$ as the inclusive probability. We
conclude that the threshold amplitude~(\ref{eq:5}) should be
interpreted with care at  large $n$ due to non-commutative nature of
large-volume and large-multiplicity limits.

\begin{sloppy}

The above finite-volume picture resembles well-known result in quantum
mechanics. Namely, consider one-dimensional particle in the potential
${V_{QM}(x)  = m_{QM}^2 x^2/2 +    \lambda_{QM} x^4/4}$. Its
transition from the ground state to the $n$-th energy level occurs with the
``probability''~\cite{Voloshin:1990mz, Cornwall:1993rh}  
\begin{equation}
  \label{eq:33}
  {\cal P}_{n}^{(QM)} = |\langle n | \hat{\cal O}| 0\rangle|^2 \sim 
  \exp\left\{ -\pi n + O(n^{1/3} \lambda_{QM}^{-2/3}) \right\}
  \qquad \mbox{at} \qquad n \gg O(\lambda_{QM}^{-1})\,, 
\end{equation}
where the prefactor is ignored and we assume that the operator
$\hat{\cal O}$ does not depend
on~$n$. Amusingly, the asymptotic formula~(\ref{eq:33}) 
does not involve the parameters~$\lambda_{QM}$ and $m_{QM}$ of the
potential. In this regard, it bears resemblance with our result
  for the multiparticle
probability ${{\cal P}_n \sim \exp\{n f_{\infty} + 
g_{\infty}/\lambda\}}$ which is dominated at $\lambda n\gg 1$ 
  by the $\lambda$-independent factor $\exp\{n
f_{\infty}\}$. Moreover, as we argued above, ${\cal P}_n$  at large $n$ can
be interpreted as the probability of transition to one of the few   
accessible $n$-particle states in a large finite box. This makes the
analogy even stronger. But there are significant differences. In
  field theory, the slope~${f_{\infty} = f_\infty(\varepsilon)}$ of the
exponent depends on energy and the subdominant term
  $g_{\infty}(\varepsilon)/\lambda \sim O(n^{0}/\lambda)$ is different.

\end{sloppy}

Forth, let us reproduce powerful
argument~\cite{Zakharov:1991rp, Libanov:1997nt, Demidov:2015bua} 
suggesting exponential suppression of multiparticle probabilities at
arbitrary values of parameters:~${F < 0}$ at any~${n\gg 1}$ and
$E$. Couple the scalar theory to the massless external fermions 
via Yukawa interaction~$y \phi \bar{\psi} \psi$ with tiny
coupling~$y$. Then dispersion
  relation and optical theorem express the amputated Green's
function~$\Pi(Q^2)$ of two $\phi$-operators in terms of the total 
fermion annihilation cross section~$\sigma_{\mathrm{tot}}(E)$: $\psi\bar{\psi} \to \mbox{anything}$~\cite{Zakharov:1991rp},
\begin{equation}
  \label{eq:34}
  \frac{d^2}{(dQ^2)^2} \, \Pi(Q^2) \Big|_{Q^2 = 0} =
  -\frac{8}{\pi y^2}\int
  \frac{dE}{E}\left(1-\frac{m^2}{E^2}\right)^2 \sigma_{\mathrm{tot}}
  (E) + O(y^2)\,. 
\end{equation}
Here, the integral in the right-hand side
converges because the physical cross section is related to the probability
and cannot grow fast with energy. Now, recall that  the standard perturbation
theory reliably calculates the two-point function
at low Euclidean momenta, and 
all nonperturbative corrections are suppressed as $\exp(- 
\mbox{const}/\lambda)$. This means that the contributions of the
 multiparticle intermediate states are also exponentially small, as
well as the $\psi \bar{\psi} \to n$ cross sections~${\sigma_n
  \leq \sigma_{\mathrm{tot}}}$ in  the right-hand side. We arrive
to the conclusion that~${F \sim \lambda \ln \sigma_n < 0}$ at arbitrary
$n\gg 1$ and~$E$, which is hard to avoid. For example, nihilistic approach~\cite{Khoze:2018qhz, 
  Belyaev:2018mtd} of dismissing the dispersion relations altogether 
barely helps: the theory cease to be sane if the sums over the intermediate
states diverge.

Fifth and finally, a notable application of our results exploits
the saddle-point 
solutions themselves. With the proper numerical input, we can
establish their reliable properties  and form the basis for future
half-analytic studies. In particular, Refs.~\cite{Khoze:2017tjt,
  Khoze:2017ifq,   Khoze:2018mey}  derived the controversial
formula~\eqref{eq:1} for the ``Higgsplosion'' scenario using
a set of assumptions on the semiclassical 
configurations at~$\lambda n \gg   1$. We can confirm one conjecture:
at large multiplicities the energy densities of our numerical
solutions form relatively narrow spherical shells of width~${\Delta
r\propto (\lambda n)^0}$ that travel inside parametrically large
``interaction'' regions~${r \lesssim O(\lambda n)^{1/2}}$, see
Sec.~\ref{sec:saddle-point-conf}, Fig.~\ref{fig:en_density}, and
  Eq.~\eqref{eq:25}. In 
  the linear regions $t\to +\infty$  our solutions satisfy even
  simpler scaling~${\phi_{\mathrm{cl}} \propto \sqrt{\lambda n}}$, see
  Eq.~\eqref{eq:4.7}.  This supports the ``thin-wall'' approach of
Refs.~\cite{Gorsky:1993ix, 
  Khoze:2017ifq}. 

\begin{figure} 
  \centerline{\includegraphics{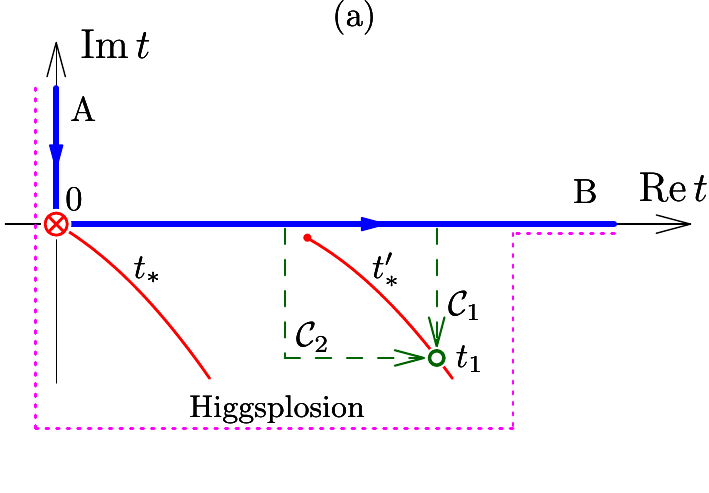}\hspace{5mm}
    \includegraphics{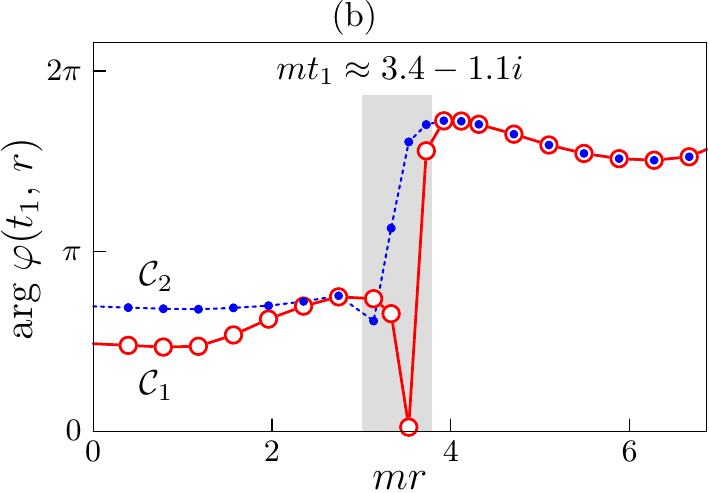}}
  \caption{(a)~Contours and singularity surfaces in the complex time
    plane (not to scale). (b)~Complex phase of the numerical
    solution~$\phi_{\mathrm{cl}}(t_1,\, r)$ as a function of~$r$ at
    the point $t = t_1$ of the second chain singularity: $t_*'(r_1) =
    t_1$. The vicinity of $r_1$ is indicated by vertical shaded
    strip. The two graphs (line-points) are  obtained using analytic
    continuations along the contours ${\cal C}_1$ and~${\cal C}_2$ in
    Fig.~(a). The solution has the following parameters: $\lambda n =
      2.51$, $\varepsilon = m/2$, $j_0 = 0.053\, m^2$, and~${\sigma =
      0.29\, m^{-1}}$.}  
  \label{fig:cut}
\end{figure}

On the other hand, we observe that the analytic structure of
our semiclassical solutions is different from the one assumed in 
Refs.~\cite{Khoze:2017tjt, Khoze:2017ifq, Khoze:2018mey}. An
important step of the latter calculation is a deformation of 
the time contour to the lower half-plane, see the dotted
(Higgsplosion) line in Fig.~\ref{fig:cut}a. That would be
legitimate if the singularities of solutions
were not crossed on the way, or if they were the poles and
their contributions could be added back to the
exponent~$F$. But in fact, all our computed saddle-point configurations have
infinite chains of 
singularities~$t_{*}(\boldsymbol{x})$,~$t_{*}'(\boldsymbol{x})$,
etc.,\ below the real time axis (at ${\rm Re}\,t>0$), as is
already   clear from the simplest 
solution~\eqref{eq:9} at~${E = n = 0}$. In Fig.~\ref{fig_entropy}b we
visualized the analytic structure of a particular numerical configuration, see also the
movie~[\citenum{movie}(b)]. Starting from the real time  
axis where this solution was originally found, we analytically continued
it to $\mathrm{Im}\, t < 0$ and $\mathrm{Im}\, t > 0$ using the
field equation. Then we displayed the complex phase of
$\phi_{\mathrm{cl}}(t,\, 0)$ (color in the figure) and marked the 
singularities~--- values of~$t$ corresponding to large
$|\phi_{\mathrm{cl}}(t,\,r)|$ at some~$r$~---  by black solid
lines. Three of them are clearly visible in
Fig.~\ref{fig_entropy}b.   

The problem is that these singularities are not  poles. In
Appendix~\ref{app:lim_j_to_0} we show that the general solution of the
field equation~\eqref{eq:2.4} has the following structure in the vicinity
of any singularity:
\begin{equation}
  \label{eq:28b}
  \phi_{\mathrm{cl}}  = \frac{C_{-1}(\boldsymbol{x})}{\tau} +
  C_0(\boldsymbol{x}) + C_1(\boldsymbol{x}) \, \tau + C_2(\boldsymbol{x}) \, \tau^2 +
  \left[B(\boldsymbol{x}) - \frac15 R_3(\boldsymbol{x}) \ln (m\tau)
    \right] \, \tau^3   + \dots \,.
\end{equation}
Here $\tau = it - it_*'(\boldsymbol{x})$ is the Euclidean time
interval to, e.g., the second singularity surface~$t_*'(\boldsymbol{x})$, the
functions  $t_*'(\boldsymbol{x})$ and $B(\boldsymbol{x})$ are
the arbitrary Cauchy data, and  all other coefficients~$C_i$
and~$R_3$ are expressed in their terms via the field
equation. Importantly, the last term with~${R_3 \ne 0}$ includes
  the logarithm 
  $\ln(m\tau)$ with a branch cut. Moreover, one can show that higher
powers of $\ln (m\tau)$ appear in higher 
orders. This means that the jumps at the branch cuts are nontrivial
and the surfaces $t_*$, $t'_*$, etc.,\ are the essential
singularities. The latter feature is a benchmark
property of
nonintegrable theories~\cite{Tabor} which distinguishes them from
exactly solvable  
cases. We checked it by numerically continuing one of the
  solutions to the two
sides of the branch cut at the singularity point $t = t_1$,
  where $t_*'(r_1) = t_1$; see the contours~${\cal C}_1$
and~${\cal C}_2$ in Fig.~\ref{fig:cut}a. Complex phases of the two
resulting configurations $\phi_{\mathrm{cl}}(t_1,\, r)$ are shown
by line-points in
Fig.~\ref{fig:cut}b, while  gray vertical strip marks the region
  near the singularity~${r  \approx r_1}$. We see that although the  two graphs
coincide at large $r$, they are different at $r<r_1$ in compliance
with the existing branch cut. 

It is worth noting that incorrect choice of the solution branch is very
dangerous, as it may give unphysical results for the
probability. Thus, it is worth applying our reliable 
numerical method to the $\phi^4$ theory 
with spontaneously broken symmetry. This will 
provide a direct test of Eq.~(\ref{eq:1}) and give valuable
information on the structure of the respective semiclassical
solutions.

Recall that our semiclassical solutions with
  $\varepsilon  \gg m$ are not sensitive to the operator~$m^2 \phi^2$
  in the Lagrangian. This suggests that our ultrarelativistic results may be relevant
  to the spontaneously broken case as well. If this is indeed the
  case, the multiparticle probability in the broken
  $\lambda \phi^4$ theory may be given  at high~$\varepsilon$ and 
  large~$\lambda n$ by our expression ${\cal P}_n \sim 
  \exp\{f_{\infty,\, \max}\, n\} \sim \exp\{-2.57 \times n\}$. As the
  probability
  grows with energy, we obtain an upper bound ${\cal
    P}_n(\varepsilon) \leq {\cal P}_n(\infty)$ at arbitrary
  $\varepsilon$ in the broken case.

From the general perspective, we believe that our numerical approach is 
scalable and can be used to describe multiparticle processes in other
bosonic field theories. It may be even helpful in distinct,
but conceptually similar situations, like semiclassical calculations
at large U(1) charge~\cite{Levkov:2017paj, Badel:2019oxl}.

\acknowledgments
This work is supported by the RFBR grant \textnumero~20-32-90013. 
Numerical calculations were performed on the Computational cluster of 
the Theoretical Division of INR RAS.


\appendix
\section{Lattice formulation}
\label{app:num_methods}
Let us describe details of the numerical method. It is convenient to
rescale the spherically-symmetric field as $\phi(t,\, r) = \chi(t,\,
  r)/r$. The action~\eqref{eq:11} takes the form,
\begin{equation}
  \label{eq:3.2}
  \frac{\lambda S_J}{2\pi} = \int dt  \int\limits_{0}^{\infty}dr
  \left[ (\partial_t \chi)^2 - ( \partial_r \chi)^2 - \chi^2
    -\frac{\chi^4}{2r^2}\right]
  +\int\limits_{0}^{\infty}dr \left[ 2irJ \chi\big|_{t=0} -
    \chi\partial_t \chi\big|_{t_{N_t+1}} \right] \!,
\end{equation}
where $m=1$, we integrated by parts and ignored boundary
terms at $r \to +\infty$ and ${t \to +i\infty}$: the field
vanishes exponentially in these regions. We have kept, however, the
boundary term at the final time $t = t_{N_t+1}$.  

\begin{sloppy}

  We use uniform spatial lattice. It has~${N_r = 256}$ sites $r_i
    \equiv  i\cdot \Delta r$ indexed by an integer~${0 \leq i \leq
      N_r-1}$. The sites fill the spherical spatial box of size~${R = 6.5 \div
      100}$. Spacing between them equals~${\Delta r = R  / (N_r -
      1)}$. On  the other hand, our temporal lattice~$\{t_j\}$ is essentially
    inhomogeneous  with steps
\begin{equation}
\label{eq:a.4}
\big| \Delta t_j \big| \equiv \big|t_{j+1} - t_j\big| = \Delta t' \left\{ 1 +
\alpha \tanh \big[ \beta (j-N_0-N_1)\big] - 
\alpha \tanh \big[ \beta (j-N_0+N_1)\big] \right\}\,,
\end{equation}
where $|\Delta t_j| \approx \Delta t'$ at infinity, the site $N_0$
  corresponds to the moment~${t_{N_0} = 0}$ when the source acts, and
the lattice is denser at~${N_0 - N_1
  \lesssim j \lesssim N_0   + N_1}$ in a way
controlled by the parameters  $\alpha$ and   
$\beta$. Recall that $t_j$  cover the complex time
contour\footnote{As we explained in Footnote~\ref{fn:1}, in
  certain cases the contour at~${j > N_0}$ is deformed into the dotted 
  line~0B$^\prime$B in Fig.~\ref{fig_cont}b. Then the complex phase of
  $\Delta t_j$ is determined by the contour tilt.} A0B in
Fig.~\ref{fig_cont}b. Hence,~$\Delta t_j = -i|\Delta t_j|$ are purely
imaginary at~${j < N_0}$ and real at larger~$j$. In practice, we
choose sufficiently small~$\Delta t'$ to resolve the out-waves
  and  make the steps near the source~${|\Delta t_{N_0}| = \Delta t'
    \big[1-2\alpha \tanh ( \beta N_1)\big]}$ approximately two orders
  of magnitude  smaller. We specify the  
  region of good resolution with values of~$N_1$~and~$\beta$ and then
  tune~$\alpha$ to the value ${\alpha\tanh ( \beta  N_1) -1/2
    \sim  10^{-2}}$. Once the lattice parameters are chosen, we
compute the sites $t_{j}$ and integers  $N_0$, $N_t$ by integrating
Eq.~\eqref{eq:a.4}. Our typical lattices have sizes in the
range~${7061\leq N_t \leq   12201}$, indices $N_0 \sim
N_t/3$ of the $t=0$ site, denser regions with~${N_1 \sim N_t/4}$\di{,} and~$\beta^{-1} \sim
10^3$, and deformation coefficients~$\alpha \approx 0.57$.

\end{sloppy}

We discretize the action using the field values $\chi_{j,\, i} =
r_i\phi_{j,\, i}$ at the lattice sites and the second-order finite-difference 
expressions
\begin{equation}
  \label{eq:14}
  \partial_t \chi \to  (\chi_{j+1,\, i} - \chi_{j,\, i}) / \Delta t_j \,,
  \qquad
   \partial_r \chi \to  (\chi_{j,\, i+1} - \chi_{j,\, i}) / \Delta r\,,
   \qquad \Delta t_j \equiv t_{j+1} - t_j\;.
\end{equation}
These derivatives are associated with the centers $(j+1/2,\, i)$ and~$(j,\, 
i+1/2)$ of time and space links, respectively. We also replace the
integrals in Eq.~(\ref{eq:3.2}) by the trapezoidal sums. Say,
\begin{equation}
  \label{eq:15}
  \int dt \, L(t) \;\;\;\; \to \;\;\;\; \sum_{j=-1}^{N_t} \Delta t_j \, L_{j+1/2} \;\;\;\;
  \mbox{or} \;\;\;\; \sum_{j=-1}^{N_t+1}\Delta \bar{t}_j\,  L_j\,,
\end{equation}
where the first expression is used for the kinetic term $L =
(\partial_t \chi)^2$ and the second one~---
for all the other terms. In Eq.~(\ref{eq:15}) we introduced 
$\Delta \bar{t}_j = (\Delta t_{j-1}+ \Delta t_j)/2$ at the inner
lattice sites and ${\Delta \bar{t}_{-1} = \Delta t_{-1}/2}$ and $\Delta 
\bar{t}_{N_t+1} = \Delta t_{N_t}/2$ at the boundaries. Discretization of the radial
integrals is performed in a similar way, but with the uniform spacing
$\Delta r$, steps $\Delta \bar{r}_i =  \Delta r$ at the inner sites,
and~$\Delta \bar{r}_{0} = \Delta \bar{r}_{N_r-1} = \Delta r/2$.

\begin{sloppy}

We impose Neumann condition $\partial_r \phi = 0$ at the
spatial boundary~${r = R}$ and require regularity
of~$\phi$ at the origin~$r=0$. In terms of~$\chi(t,\, r)$
this reads,
\begin{equation}
  \label{eq:17}
  \partial_r \chi(t,\, R) = R^{-1}\, \chi(t,\, R) \qquad
  \qquad\mbox{and} \qquad\qquad  \chi(t,\, 0) = 0\,.
\end{equation}
To put such conditions on the same footing with the field
equation, we add
the term~${\lambda \Delta S_J / 2\pi = \int dt \, \left[ \chi^2 (t,\,
    R)/R  - 2\chi(t,\, 0) \partial_r \chi(t,\,0) \right]}$ to the classical
action~(\ref{eq:3.2}). After that Eqs.~(\ref{eq:17}) are obtained by 
extremizing $S_J$ with respect to~$\chi (t,\, R)$ and~$\chi(t,\,
0)$. This modification of the action does not change the value of the
suppression exponent because the saddle-point value
  of~$\chi_\mathrm{cl}(t,\, R)$ is exponentially small
  and~$\chi_{\mathrm{cl}}(t,\, 0)$ is zero. We discretize the extra
term~$\Delta S_J$ in the same way as the others. 

\end{sloppy}

Substitutions~(\ref{eq:14}) and (\ref{eq:15}) give  the lattice
action, 
\begin{align}
  \frac{\lambda S_J}{2\pi}  & =
    \sum_{j=-1}^{N_t} \sum_{i=0}^{N_r-1} \frac{\Delta
    \bar{r}_i}{\Delta t_j} \, (\chi_{j+1,\, i} - \chi_{j,\, i})^2
    - \sum_{j=-1}^{N_t+1} \sum_{i=0}^{N_r-2} \frac{\Delta
      \bar{t}_j}{\Delta r} \,(\chi_{j,\, i+1}-\chi_{j,\, i})^2
    + 2i \sum_{i=0}^{N_r-1} \Delta \bar{r}_i \, r_i
      J_{i} \, \chi_{N_0,\, i}
    \nonumber \\
    & - \sum_{j=-1}^{N_t+1} \sum_{i=0}^{N_r-1} \Delta \bar{t}_j  \Delta
    \bar{r}_i \left( \chi^2_{j,\, i}  + \frac{\chi^4_{j,\, i}}{2
      r_i^2} \right)
    + \sum_{j=-1}^{N_t+1} \Delta \bar{t}_j\, \left( \frac{\chi^2_{j,\, N_r-1}}{R}
    - \frac{2}{\Delta r} \chi_{j,\, 0} \chi_{j,\, 1} \right)
    \label{eq:a.1}
    \\
    & - \int\limits_{0}^{\infty}dr \, 
    \chi\partial_t \chi\big|_{t_{N_t+1}}\,,
    \nonumber
\end{align}
where $J_i \equiv J(r_i)$ and the boundary term at $t =
  t_{N_t+1}$  is still written in the continuous form: we will 
discretize it later. Lattice field equations and boundary
  conditions at $r = 0$,  $R$ are obtained by extremizing the first
  two lines of Eq.~(\ref{eq:a.1}) with respect to $\chi_{j,\,   i}$ at
  $0 \leq j \leq N_t$ and ignoring the last boundary term; see
  Eq.~\eqref{eq:12}. 

Next, we derive finite-difference  boundary conditions in the asymptotic past and
future. Equation~\eqref{eq:2.5} can
be imposed at  the very first  time site,
\begin{equation}
  \label{eq:16}
  \chi_{-1,\, i} = 0\,.
\end{equation}
At  large positive times, the evolution is linear and the time
lattice is almost uniform: ${\Delta t_j \approx \Delta
t'}$. The lattice field equation simplifies,
\begin{equation}
  \label{eq:19}
  \frac{\chi_{j+1,\, i} + \chi_{j-1,\, i} - 2\chi_{j,\, i}}{\Delta
  t'^2} - \sum_{i'}\Delta_{i,\, i'} \chi_{j, \, i'} + \chi_{j,\, i} = 0
  \qquad \mbox{at large $t_j$}\,,
\end{equation}
where the three-diagonal matrix
$\Delta_{i,\, i'}$ replacing the 
Laplacian can be explicitly deduced from the above 
action. We solve  
Eq.~(\ref{eq:19}) using the basis of eigenvectors~$\psi^{(l)}_i$ 
diagonalizing the matrix: ${\Delta\psi^{(l)} = -k_l^2\, \psi^{(l)}{\equiv
-\left(\omega_l^2-1\right)\, \psi^{(l)}}}$, 
where $k_l$ are the discrete analogs of momenta and~$l$ is 
an integer. The solution
\begin{equation}
\label{eq:a.10}
\chi_{j,\, i}= \sum_{l=1}^{N_r}
\frac{\psi_i^{(l)}}{{\sqrt{2\omega_l}}}\; \left(\,  a_l \,
\mathrm{e}^{-i{\mu_l} t_j}+ b_l^* \, \mathrm{e}^{i{\mu_l}
  t_j} \right)\;,
\qquad 
{\mu_l} = \frac{2}{\Delta t'} \arcsin \left\{ \frac{\Delta t'}{2}
  {\omega_l}\right\}
\end{equation}
is parametrized by arbitrary complex amplitudes $a_l$ and~$b_l^*$
of  terms oscillating with lattice frequencies $\pm \mu_l$.  In the
continuous limit, the eigenvectors turn into spherical
  harmonics, ${\psi_i^{(l)} \propto  \sin(k_l r_i)}$, the
    frequencies start to obey the standard dispersion relation $\mu_l^2 \to
k_l^2  + 1$, and~$a_l$ and~$b_l$ become 
  proportional to the negative- and positive-frequency amplitudes.
This discloses Eq.~\eqref{eq:a.10} as a discrete version of the free
wave decomposition~\eqref{eq:2.6}. In practice, we numerically compute
the spectrum~$\{ \psi^{(l)},\, k_l^2{\equiv \omega_l^2 - 1}\}$ of~$\Delta_{i,\,  i'}$
and extract   
the amplitudes from the field values at the two last time sites:
\begin{align}
\label{eq:a.11}
& a_l= \frac{i \mathrm{e}^{ i{\mu_l} 
      t_{N_t+1}} \sqrt{{\omega_l}}}{\sqrt{2}\, \sin({\mu_l} \Delta 
  t')}  \;  \sum_{i=0}^{N_r-1} \Delta \bar{r}_i \, \psi_i^{(l)} 
 \left( \chi_{N_t+1, \, i} \; \mathrm{e}^{- i{\mu_l} \Delta
    t'}-\chi_{N_t,\, i} \right)\,,\\[1ex] \nonumber
 &b_l^*= \frac{\mathrm{e}^{-i{\mu_l}  t_{N_t+1}} \sqrt{{\omega_l}}}{ i \sqrt{2} \sin({\mu_l} \Delta t')} 
 \;\sum_{i=0}^{N_r-1} \Delta \bar{r}_i \, \psi_i^{(l)} \left(
 \chi_{N_t+1,\, i}\; \mathrm{e}^{i{\mu_l} \Delta 
  t'}-\chi_{N_t,\, i} \right)\,,
\end{align}
where normalization\footnote{This scalar product is related to the fact
  that the matrix $\Delta \bar{r}_i\, \Delta_{i,\, i'}$ is
  symmetric, while
  $\Delta_{i,\, i'}$ itself is not, cf.\ Eqs.~\eqref{eq:a.1} and
  \eqref{eq:19}.} $\sum_i \Delta \bar{r}_i \, \psi_i^{(l)}
\psi_i^{(l')} = \delta_{ll'}$ 
is assumed. The final boundary 
conditions~\eqref{eq:2.7} in the discrete form\footnote{One
  eigenvector of $\Delta_{i,\,  i'}$ has the form
  $\psi^{(N_r)}_i \propto \delta_{0,i}$ due to our choice of the boundary 
  term at~${i = 0}$ in Eq.~\eqref{eq:a.1}. In this case
  Eq.~\eqref{eq:20} together with the boundary condition
  $\chi_{j,\, 0} = 0$ ascertain that $a_{N_r}=b_{N_r}\equiv 0$ for all
  our solutions.}  
\begin{equation}
  \label{eq:20}
  a_l = {\rm e}^{-\theta + 2\omega_{l}T}\, b_{l}
\end{equation}
impose a set of linear relations on $\chi_{N_t,\,  i}$ and $\chi_{N_t
  +1,\, i}$.

The last term in the action~\eqref{eq:a.1} is
discretized by replacing
\begin{equation}
  \label{eq:18}
  \int\limits_{0}^{\infty}dr \, \chi\partial_t
  \chi\big|_{t_{N_t+1}} \to \sum_{i=0}^{N_r-1} \sum_{l=1}^{N_r} \Delta
  \bar{r}_i \, \chi_{N_t+1, \, i} \, \psi_{i}^{(l)} \left(
  \frac{\tilde{\chi}_{N_t+1,\, l} - \tilde{\chi}_{N_t,\, l}}{\Delta t_{N_t}} - 
  \frac{\Delta t_{N_t}}{2} \, \omega_l^2 \, \tilde{\chi}_{N_t+1,\, l} \right) \,,
\end{equation}
where $\tilde{\chi}_{j,\, l} \equiv \sum_{i} \Delta \bar{r}_i \, \psi_i^{(l)} \,
\chi_{j,\,i}$ is the field in the basis of free waves on the
lattice. One can check that Eq.~(\ref{eq:18}) is the second-order
  discretization in
$\Delta t_j$ by performing Taylor series expansion in this
parameter and using Eq.~(\ref{eq:a.10}).

Given the representation~\eqref{eq:a.10}, we immediately write
parameters of the final state as
\begin{equation}
\label{eq:a.13}
\lambda E =  4\pi \sum_l \omega_l a_l b_l^* \,, \qquad \qquad 
\lambda n = 4\pi \sum_l a_l b_l^* \,.
\end{equation}
It is straightforward to see that these expressions reproduce
Eqs.~\eqref{eq:2.8} in the continuous limit. In particular, the
continuous occupation numbers equal
  \begin{equation}
    a_{\bf k}b_{\bf k}^* \approx \frac{a_lb_l^*}{k_l^2\Delta k_l}\,,
  \end{equation}
  where $\Delta k_l=k_l-k_{l-1}\approx \pi/R$. We exploit this
    matching to plot Fig.~\ref{fig_n_t_limit}.

Discretized field equation~\eqref{eq:12} with the boundary conditions
(\ref{eq:16}), (\ref{eq:20}) and expressions~\eqref{eq:a.13} for
$\lambda E$ and $\lambda n$ form an algebraic
system of nonlinear equations $G_{\alpha} = 0$ for the unknowns ${y_\alpha
  \equiv \{\phi_{j,\, i},\, T,\, \theta\}}$. We solve this system
as described in the main text. After finding the solution, we compute the
suppression exponent $F_J$ using Eqs.~\eqref{eq:2.9a},
  \eqref{eq:a.1}, and~\eqref{eq:18}.

It is worth noting that the nonlinear energy~\eqref{eq:7}
is discretized in the same way as the classical action. We check
its conservation along the parts A0 and 0B of the time contour to
determine the discretization errors. Also, difference between the exact
and free-wave energies, Eqs.~\eqref{eq:7}  and~\eqref{eq:a.13},
estimates nonlinear effects in the final state which should be small.

\section{Solutions in the linear theory}
\label{sec:semicl-solut-line}

\begin{sloppy}

In this Appendix we perform semiclassical calculations in the 
free theory with a source. Start with the
solution~\eqref{eq:3.5}. Evaluating  the integral over $k^0$ in its    
second term, we get, 
\begin{align} 
  \label{eq:21}
  & \phi_{\mathrm{cl}}^{(\mathrm{lin})} = - \int \frac{d^3
    \boldsymbol{k}}{(2\pi)^3}\,
  \frac{\mathrm{e}^{i\omega_{\boldsymbol{k}}t - i\boldsymbol{k}
      \boldsymbol{x}}}{2\omega_{\boldsymbol{k}}} \left[ \,
    J(-\boldsymbol{k}) + J^*(\boldsymbol{k}) \, \mathrm{e}^{\theta -
      2\omega_{\boldsymbol{k}} T}\, \right] & \mbox{at} \quad
  t<0\,,\\ \label{eq:22}
  & \phi_{\mathrm{cl}}^{(\mathrm{lin})} = - \int \frac{d^3
    \boldsymbol{k}}{(2\pi)^3}\,
  \frac{\mathrm{e}^{i\boldsymbol{k}
      \boldsymbol{x}}}{2\omega_{\boldsymbol{k}}}
  \left[ \, J(\boldsymbol{k})\,\mathrm{e}^{-i\omega_{\boldsymbol{k}} t}
    + J^*(-\boldsymbol{k}) \, \mathrm{e}^{i \omega_{\boldsymbol{k}}t +
      \theta - 2\omega_{\boldsymbol{k}} T}\, \right] & \mbox{at} \quad t>0\,.
\end{align}
We analytically continue this function from $t<0$ to the upper half of
  imaginary axis~${t = i|t|}$. Clearly, it decreases as $t \to
  +i\infty$ in accordance with Eq.~\eqref{eq:2.5}. At $t>0$, the
configuration~\eqref{eq:22} explicitly 
satisfies the boundary conditions
in the infinite future~(\ref{eq:2.6}), (\ref{eq:2.7}) with
${a_{\boldsymbol{k}} = -J(\boldsymbol{k}) / 
\sqrt{2\omega_{\boldsymbol{k}} 
  (2\pi)^{3}}}$ and ${b_{\boldsymbol{k}} = - J(\boldsymbol{k}) \,
  \mathrm{e}^{\theta -   2\omega_{\boldsymbol{k}} T} /
  \sqrt{2\omega_{\boldsymbol{k}} (2\pi)^{3}}}$. 

\end{sloppy}

Given $a_{\boldsymbol{k}}$ and $b_{\boldsymbol{k}}$, we evaluate the
out-state parameters~(\ref{eq:2.8}) as
\begin{equation}
  \label{eq:23}
  \lambda E = \mathrm{e}^{\theta}\int \frac{d^3 \boldsymbol{k}}{2 (2\pi)^3} \,
  |J(\boldsymbol{k})|^2 \,
  \mathrm{e}^{-2\omega_{\boldsymbol{k}} T}\, \qquad 
  \lambda n  = \mathrm{e}^{\theta}\int \frac{d^3
    \boldsymbol{k}}{2\omega_{\boldsymbol{k}} (2\pi)^3} \, 
  |J(\boldsymbol{k})|^2 \,
  \mathrm{e}^{-2\omega_{\boldsymbol{k}} T}\,.
\end{equation}
One can express $\theta$ from the second of these equations. Then the
first relates $T$ to $\varepsilon = E/n-m$. Finally, the exponent
\begin{equation}
\label{eq:3.8}
F_J = 2\lambda ET-\lambda n \theta + \mathrm{Re}\int \frac{d^3 \boldsymbol{k}
}{(2\pi)^3} \, 
\frac{J(\boldsymbol{k}) }{2\omega_{\boldsymbol{k}}} \,
\left[ J(-\boldsymbol{k})\,  + J^*(\boldsymbol{k}) \,
  e^{\theta-2\omega_{\mathbf{k}}T} \right]
\end{equation}
 is obtained by substituting Eq.~(\ref{eq:3.5}) into
 Eqs.~(\ref{eq:2.1}) and~(\ref{eq:2.9a}) and ignoring the
 interaction term.  

We use the above expressions in the following way. For given
$\lambda n$ and $\varepsilon$, we express $\theta$ and get a
  nonlinear equation for $T = T(\varepsilon)$ from
Eqs.~(\ref{eq:23}). The latter equation is solved by
 binary search~\cite{NR} and  
numerical computation of the $\boldsymbol{k}$ integrals. Recall that
we always exploit the Gaussian source~\eqref{eq:2.3} with the
Fourier image 
\begin{equation} 
\label{eq:3.7}
J(\boldsymbol{k}) \equiv \int d^3 \boldsymbol{x} \, J(\boldsymbol{x})
\, {\rm e}^{- i\boldsymbol{kx}} = j_0 \;(2\pi \sigma^2)^{3/2}\;
{\rm e}^{-\boldsymbol{k}^2\sigma^2 / 2}\,.
\end{equation}
We determine the starting configuration for the main numerical
procedure of this paper by computing the $\boldsymbol{k}$ integrals in
Eqs.~(\ref{eq:21}), (\ref{eq:22})  at every lattice point $(t_j,\, r_i = 
|\boldsymbol{x}|_i)$. The exponent in the linear theory is given by
Eq.~(\ref{eq:3.8}). This result for~$F_J$  is displayed with the dotted
line in Fig.~\ref{fig_f_j}a.


\section{Singularity structure and the limit $J \to 0$}
\label{app:lim_j_to_0}

Let us analyze the structure of semiclassical solutions near
their singularities~$t = 
t_*(\boldsymbol{x})$. This will establish their~$j_0$ dependence and allow us
to perform extrapolation $j_0 \to 0$. Also, we will be able to prove that the ``main''
singularity surface touches the physical contour at~$j_0 = 0$. 

We introduce Euclidean time interval to the singularity,
\begin{equation}
\label{eq:c.1}
\tau = i \, [t - t_*(\boldsymbol{x})]\,,
\end{equation}
and change the coordinates in the field
equation~\eqref{eq:2.4} to $\tau$ and~$\boldsymbol{x}$. This gives:
\begin{equation}
\label{eq:c.2}
-\left[ 1 - (\partial_k t_*)^2 \right] \partial_\tau^2 \phi_{\mathrm{cl}} + 
2i \partial_k t_* \,  \partial_k \partial_\tau \phi_{\mathrm{cl}} +
i \Delta t_*  \, \partial_\tau \phi_{\mathrm{cl}}
- \Delta \phi_{\mathrm{cl}} + m^2 \phi_{\mathrm{cl}} + \phi_{\mathrm{cl}}^3 = 0\,,
\end{equation}
where~$k=1,2,3$ indexes spatial coordinates, $\Delta$ is the Laplacian, and 
we so far ignore the source in the right-hand side of Eq.~\eqref{eq:2.4}. In the
vicinity of the singularity, it is natural to use the power series
in~$\tau$:
\begin{equation}
\label{eq:c.3}
\phi_{\mathrm{cl}} (\tau, \, \boldsymbol{x}) = \sum_{n = -1}^{+\infty} C_n
(\tau,\, \boldsymbol{x}) \; \tau^n \,,
\end{equation}
where the first~$\tau^{-1}$ term is motivated by the
solution~\eqref{eq:9}. Soon we will see that the coefficients~$C_n$ of 
the above expansion are either $\tau$-independent or depend slowly
(logarithmically) on this coordinate.

We substitute Eq.~(\ref{eq:c.3}) into the field equation and solve it
order-by-order in $\tau$. The leading order gives
$C_{-1}=\sqrt{2-2(\partial_k t_*)^2}$, and in the next orders we
  obtain the following equations
\begin{equation}
\label{eq:c.4}
\left(\,-\partial_\tau^2 + 6\tau^{-2} \,\right)  ( \tau^nC_n)
= \tau^{n-2} R_n\,.
\end{equation}
Here $R_n$ depend on $t_*(\boldsymbol{x})$ as well as on the coefficients $C_m$
with lower index $m\leq n-1$; one can explicitly deduce these right-hand sides  from
Eq.~(\ref{eq:c.2}). It is clear that Eq.~(\ref{eq:c.4}) sequentially  
determines~$C_n$. In particular,
\begin{equation}
\label{eq:c.6}
C_n = -\frac{R_{n}[t_*,\, C_{n-1},\, C_{n-2},\,\dots]}{(n+2)(n-3)} \qquad
\mbox{with}\qquad n = 0,\, 1, \mbox{or 2}
\end{equation}
do not depend on~$\tau$. They can be expressed in terms of
$t_*(\boldsymbol{x})$ and its derivatives by substituting all previous
$C_{m}$ with $m\leq n-1$ into $R_n$. 

But there are two subtleties. First, the operator in the left-hand
side of Eq.~(\ref{eq:c.4}) annihilates the function
\begin{equation}
  \label{eq:2}
  \delta \phi_{\mathrm{cl}} = \frac{i C_{-1}}{\tau^2} \; \delta
  t_*(\boldsymbol{x}) + B(\boldsymbol{x})\, \tau^3
\end{equation}
for arbitrary $\delta t_*(\boldsymbol{x})$ and~$B(\boldsymbol{x})$.
This is the freedom of solving Eqs.~(\ref{eq:c.4}): we can
change the singularity surface~$t_* \to t_*(\boldsymbol{x}) + \delta
t_*(\boldsymbol{x})$ and add~$\tau$-independent part
$B(\boldsymbol{x})$ to~$C_3$. Second, the coefficient $C_3$ itself cannot  be 
$\tau$--independent, or Eq.~(\ref{eq:c.4}) would not be satisfied
at~${n=3}$. Indeed, the left-hand side of this equation equals
  zero for $C_3 = B(\boldsymbol{x})$ in disagreement with 
\begin{equation}
\label{eq:c.7}
R_3(\boldsymbol{x}) = \frac{2}{C_{-1}^3}\, \partial_k \left( -4iC_{-1}C_2 \,\partial_k t_*
  +C_{-1}\,\partial_k C_1 - C_1 \,\partial_k C_{-1} \right) \ne 0\,.
\end{equation}
In particular, $R_3(0) =  2\sqrt{2}\,[m\Delta t_*(0)/3]^2 \ne 0$ at
$\boldsymbol{x}=0$ in the spherically-symmetric case.
Solving Eq.~(\ref{eq:c.4}) at $n=3$, we obtain, 
\begin{equation}
\label{eq:c.8}
C_3(\tau,\, \boldsymbol{x}) = B (\boldsymbol{x}) - \frac15 \,
R_3(\boldsymbol{x}) \ln(m\tau) \,,
\end{equation}
where the logarithmic term is important for establishing
the analytic structure of the solution. Now, it is clear that the
  singularity~${t = t_*(\boldsymbol{x})}$ is  a 
branching point. 

One can demonstrate that at higher orders~${n\geq 4}$ the
  coefficients $C_{n}(\tau,\, \boldsymbol{x})$ include powers of $\ln(m\tau)$ coming
  from nonlinear terms in the right-hand sides~$R_n$. This tells us that~${t =
t_*(\boldsymbol{x})}$ is an essential singularity with a branch
cut. It is worth reminding that existence of such 
singularities in the general solution is a benchmark property
of nonintegrable models like our scalar $\lambda\phi^4$ theory.

To sum up, the recurrent relations (\ref{eq:c.4}) express all
$C_n(\tau,\, \boldsymbol{x})$  in terms of two arbitrary functions
$t_*(\boldsymbol{x})$ and 
$B(\boldsymbol{x})$. This is the precisely the amount of Cauchy data
required for the second-order field equation; hence, general solution
of the latter has the form~(\ref{eq:c.3}) near every singularity,
indeed. Thereby, we justified Eq.~(\ref{eq:28b}) from
the main text. Representation~(\ref{eq:c.3})  
with logarithmically dependent coefficients is known in the literature
as logarithmic~$\Psi$~series~\cite{Tabor}.

Now, let us restore the source~$J(\boldsymbol{x})$ at $t=0$ in
the right-hand side of the field equation. The  
respective solution consists of two analytic
  functions~$\phi_{\mathrm{cl}}^{-}(t,\, \boldsymbol{x})$ 
and~$\phi^{+}_{\mathrm{cl}}(t,\, \boldsymbol{x})$ defined
on the intervals~A0 and~0B of the complex time contour.
The functions are sewed at~${t=0}$ according to
Eqs.~\eqref{eq:6}. Assuming that $\phi_{\mathrm{cl}}^{\pm}$ have singularities
near~${t=\boldsymbol{x}=0}$, we can write them as the 
series~(\ref{eq:c.3}) in the vicinity of this point. We thus
  parametrize the two parts of the solution with~$t_*^\pm(\boldsymbol{x})$
  and~$B^\pm(\boldsymbol{x})$. To the leading order in~$t_*$, the
sewing conditions give,
\begin{equation}
  \label{eq:29}
  t_*^+ - t_*^-  \approx -\frac{J(\boldsymbol{x})\,
    t_*^3(\boldsymbol{x})}{5\sqrt{2}} \,, \qquad\qquad  
  B^+ - B^- \approx - \frac{J(\boldsymbol{x})}{5\,t_*^{2}(\boldsymbol{x})}\,.
\end{equation}
We see that the difference between the two singularity surfaces
$t_*^\pm \approx t_*$  is parametrically suppressed by
both $t_*^3(\boldsymbol{x})$ and $J(\boldsymbol{x})$, whereas the
jump of the parameter $B$ may be 
large~\cite{Son:1995wz}.  This justifies Eq.~\eqref{eq:24} which includes a
  single singularity surface.

Next, we extract $j_0$ dependence of  the solution using the
expression~\eqref{eq:8} for energy. Indeed, since the source 
is narrow, the integral  in that expression is saturated in the small
vicinity of ${t = \boldsymbol{x} = 0}$ where we can adopt the
approximation~\eqref{eq:24}: use the leading singular
term~${\phi_{\mathrm{cl}}^{\pm}   \approx \sqrt{2}/\tau}$ of the
solution and ignore difference between the two singularity
surfaces~${t_*^{\pm}(\boldsymbol{x}) \approx t_{*,\, 0} + 
  t_{*,\, 2}\, \boldsymbol{x}^2}$. Here
the complex parameters~$t_{*,0}$ and~${t_{*,\, 2} \equiv \Delta t_*(0)/6}$
characterize shift and  curvature of the singularity surface  at 
$\boldsymbol{x}=0$. Expression~\eqref{eq:8} takes the form,
\begin{equation}
\label{eq:31}
  \lambda E \approx -\int 
  \frac{J(\boldsymbol{x})\, d^3 \boldsymbol{x} \, \sqrt{2}}{(t_{*,\, 0} + t_{*,\, 2} 
    \,\boldsymbol{x}^2)^2} = \frac{4 \pi j_0
  \sqrt{2}}{(it_{*,0})^{1/2} (it_{*,\, 2})^{3/2}}\left[\, 
  \frac{\pi}{4}\, {\rm e}^\zeta \, (1+2\zeta) 
  \,{\rm erfc}\sqrt{\zeta}
  - \frac{\sqrt{\pi \zeta}}{2} \, \right]\, ,
\end{equation}
where in the second equality we calculated the integral for
the Gaussian source~\eqref{eq:2.3} of strength $j_0$ and width
$\sigma$. We also introduced a combination ${\zeta = t_{*, \, 0}
  / (2 t_{*,\, 2}\, \sigma^2)}$ and exploited the complementary
error function ${\rm erfc}(z)$.  

Suppose the singularity surface remains smooth in the limit $j_0 \to
0$ and $j_0 / \sigma = \mbox{const}$, i.e.~$t_{*,\, 2}$ is finite.
Then finiteness of energy~$E$ in this limit implies that $t_{*,\,
  0}  \to O(j_0^2)$ and $\zeta$ tends to a constant. This result is
used in Sec.~\ref{sec:saddle-point-conf} and
 numerically confirmed\footnote{We performed a stronger
 test. To this end we computed the singularity surfaces of the numerical solutions
   at different~$j_0 / \sigma$ 
   and small~$j_0$. The combinations~$j_0 / (t_{*,\, 0}^2t_{*,\, 2}^3)$
 and  $\zeta$ were related by Eq.~(\ref{eq:31}).} in
Fig.~\ref{fig_re_phi_tau0_j}b. It implies, in particular, that the
singularity surface  touches the point $t = \boldsymbol{x} = 0$
and the semiclassical solutions are truly singular at $j_0 = 0$.
Besides, together with Eqs.~(\ref{eq:29}) this scaling 
establishes sewing condition between the two parts of solutions at 
zero source~\cite{Son:1995wz}: their singularity surfaces should
coincide,~${t_*^{+} =  t_*^{-}}$, and the jump of
$B(\boldsymbol{x})$ should be proportional to the
$\delta$-function:~${B^+ - B^- = \lambda E \,
  \delta^{3}(\boldsymbol{x}) / 5\sqrt{2}}$. Note that the latter
condition should be taken cautiously, as powers of~$B$ are present
in  the higher-order terms of the $\Psi$ series. 

We are ready to conclude that the semiclassical solution and all its
characteristics~$F_J$,~$\theta$, and $T$ can be expressed as power
series in~$j_0^2$ at~$j_0/\sigma = \mbox{const}$.  Indeed, although
relation~(\ref{eq:31}) was derived in the leading order,
corrections to it go in powers of $t_*(\boldsymbol{x}) \propto
  j_0^2$.  Thus, the solution~$t_{*,\,   0}$ of this
  equation can be expressed as  series in $j_0^2$ at a fixed
energy. Expanding the coefficients in Eq.~(\ref{eq:c.3})~---
nonlinear  functions of  $t_*(\boldsymbol{x})$~--- in corrections,
one turns the entire solution  into series in~$j_0^2$. This fact
was used in  Sec.~\ref{sec:way_to_sing_solut} for extrapolating
results to~$j_0 = 0$, see  Eq.~\eqref{eq:3.11}.  

\bibliographystyle{JHEP}
\bibliography{n_to_infty}
\end{document}